\documentclass{aastex}

\usepackage{graphicx}

\usepackage{emulateapj5}
\usepackage{times}

\makeatletter

\newenvironment{inlinefigure}{%
\def\@captype{figure}%
\noindent\begin{minipage}{0.999\linewidth}\begin{center}}
{\end{center}\end{minipage}\smallskip}
\makeatother

\usepackage{mathptmx}

\def\lax    {{_<\atop^{\sim}}}
\def\gax    {{_>\atop^{\sim}}}
\def\Mo     {{\rm M}_{\odot}}
\def\Lo     {{\rm L}_{\odot}}

\begin{document}

\slugcomment{submitted to {\em The Astrophysical Journal}}

\title{Isotropic AGN Heating with Small Radio Quiet Bubbles in the NGC 5044 Group}

\author{Laurence P. David\footnote{Harvard-Smithsonian Center for Astrophysics, 60 Garden St.,
Cambridge, MA 02138}, Christine Jones$^a$, Simona Giacintucci$^a$, William Forman$^a$, Paul Nulsen$^a$, Jan Vrtilek$^a$, Ewan O'Sullivan$^a$ and Some Raychaudhury\footnote{School of Physics and Astronomy, University of Birmingham, Birmingham B15 2TT}}

\shorttitle{\emph NGC 5044}

\begin{abstract}

\end{abstract}

A Chandra observation of the X-ray bright group NGC 5044 shows that 
the X-ray emitting gas has been strongly perturbed 
by recent outbursts from the central AGN and also by motion of the
central dominant galaxy relative to the group gas.  The NGC 5044 group hosts many 
small radio quiet cavities with a nearly isotropic distribution, cool filaments,
a semi-circular cold front and a two-armed spiral shaped feature of cool gas.  
A GMRT observation of NGC 5044 at 610~MHz shows the presence of  
extended radio emission with a "torus-shaped" morphology.
The largest X-ray filament appears to thread the radio torus, suggesting that the 
lower entropy gas within the filament is material being uplifted from the center of the group.  
The radio emission at 235~MHz is much more extended than the emission at 610~MHz, 
with little overlap between the two frequencies. One component of
the 235~MHz emission passes through the largest X-ray cavity 
and is then deflected just behind the cold front. 
A second detached radio lobe is also detected at 235~MHz beyond the cold front.
All of the smaller X-ray cavities in the center of NGC 5044 are undetected in the 
GMRT observations. Since the smaller bubbles are probably no longer momentum 
driven by the central AGN, 
their motion will be affected by the group "weather" as they buoyantly rise outward. 
Hence, most of the enthalpy within the smaller bubbles will likely
be deposited near the group center and isotropized by the group weather.
The total mechanical power of the smaller radio quiet cavities is 
$P_c = 9.2 \times 10^{41}$erg~s$^{-1}$ which is sufficient to suppress about 
one-half of the total radiative cooling within the central 10~kpc.  This is consistent 
with the presence of H$\alpha$ emission within this region which shows that at 
least some of the gas is able to cool.  The mechanical heating power 
of the larger southern cavity, located between 10 and 20~kpc, is six times greater than 
the combined mechanical power of the smaller radio quiet cavities and could suppress
all radiative cooling within the central 25~kpc if the energy were deposited 
and isotropized within this region.  Within the central 20~kpc, emission from 
LMXBs is a significant component of the X-ray emission above 2~keV. 
The presence of hard X-ray emission from unresolved LMXBs makes it difficult to place 
strong constraints on the amount of shock heated gas within the X-ray cavities.



\section{Introduction}

Chandra and XMM-Newton observations of groups and clusters of galaxies
have led to significant changes in the cooling flow model
(e.g., McNamara et al. 2000; David et al. 2001; Fabian et al. 2003;
Blanton et al. 2003;
Peterson et al. 2003; Nulsen et al. 2005;
Peterson \& Fabian 2006; Forman et al. 2007; McNamara \& Nulsen 2007 and 
references therein).  
The primary revision to the cooling flow model has been the addition of a 
strong feedback mechanism between the central AGN and the cooling of the 
hot gas.  AGN outbursts in 
the central dominant galaxy in groups and clusters, which are themselves
fueled by the accretion of cooling gas, can produce shocks, cavities
and sound waves, all of which lead to re-heating of the cooling gas.
This AGN-cooling flow feedback mechanism is also an important process
in galaxy formation and can help explain the observed correlation between 
bulge mass (or stellar velocity dispersion) and central black hole mass 
(e.g., Gebhardt et al. 2000) and the cut-off in the number of 
massive galaxies (Croton et al. 2006).  There have been extensive 
studies with deep Chandra and XMM-Newton observations of several rich 
clusters (e.g., Perseus - Fabian et al. 2006; Centaurus - Fabian et al. 2005;
Virgo - Forman et al. 2007) that show the full complexity of AGN
induced structure in the hot gas, but far fewer detailed studies of the 
the AGN-cooling flow connection in groups of galaxies, which harbor
the majority of baryonic matter in the universe.  In addition to AGN induced 
X-ray features in groups and clusters, motion of the central dominant galaxy 
relative to the hot gas can also produce sharp fronts in the X-ray images
(see Markevitch \& Vikhlinin 2007 for a review).  In this paper, we present the 
results of a moderately deep Chandra observation of the X-ray bright NGC 5044 group of 
galaxies and discuss the variety of AGN and motion induced features evident
in the X-ray data.

The NGC 5044 group of galaxies is one of the X-ray brightest groups in the 
sky with a redshift of z=0.009.  The large scale X-ray morphology of the NGC 5044 
group, as revealed by a ROSAT PSPC observation, is very smooth and nearly spherically
symmetric (David et al. 1995).  Within the central 20~kpc, however, there were some 
indications from the PSPC observation that the group is not fully relaxed.
The PSPC image showed that the peak of the X-ray emission in the group, 
which is coincident with the optical centroid of NGC 5044, 
is off-set from the centroid of the outer X-ray contours, suggesting that 
NGC 5044 is in motion with respect to the center of the group potential. 
A "tear drop" shaped feature of cooler gas extending to the south-east 
from NGC 5044 was also detected in the PSPC image. This was interpreted as a cooling 
wake by David et al. (1995), i.e., cooling gas that was
gravitationally focused into the wake of NGC 5044. ASCA observations 
of NGC 5044 also showed evidence for multi-phase gas in the center of the group
(Buote 1999).

NGC 5044 was previously observed by Chandra for 20~ksec on Mar. 19, 2000.  A joint 
analysis of this earlier Chandra observation and a XMM-Newton observation by 
Buote et al. (2003) 

\begin{inlinefigure}
\center{\includegraphics*[width=1.00\linewidth,bb=96 239 448 555,clip]{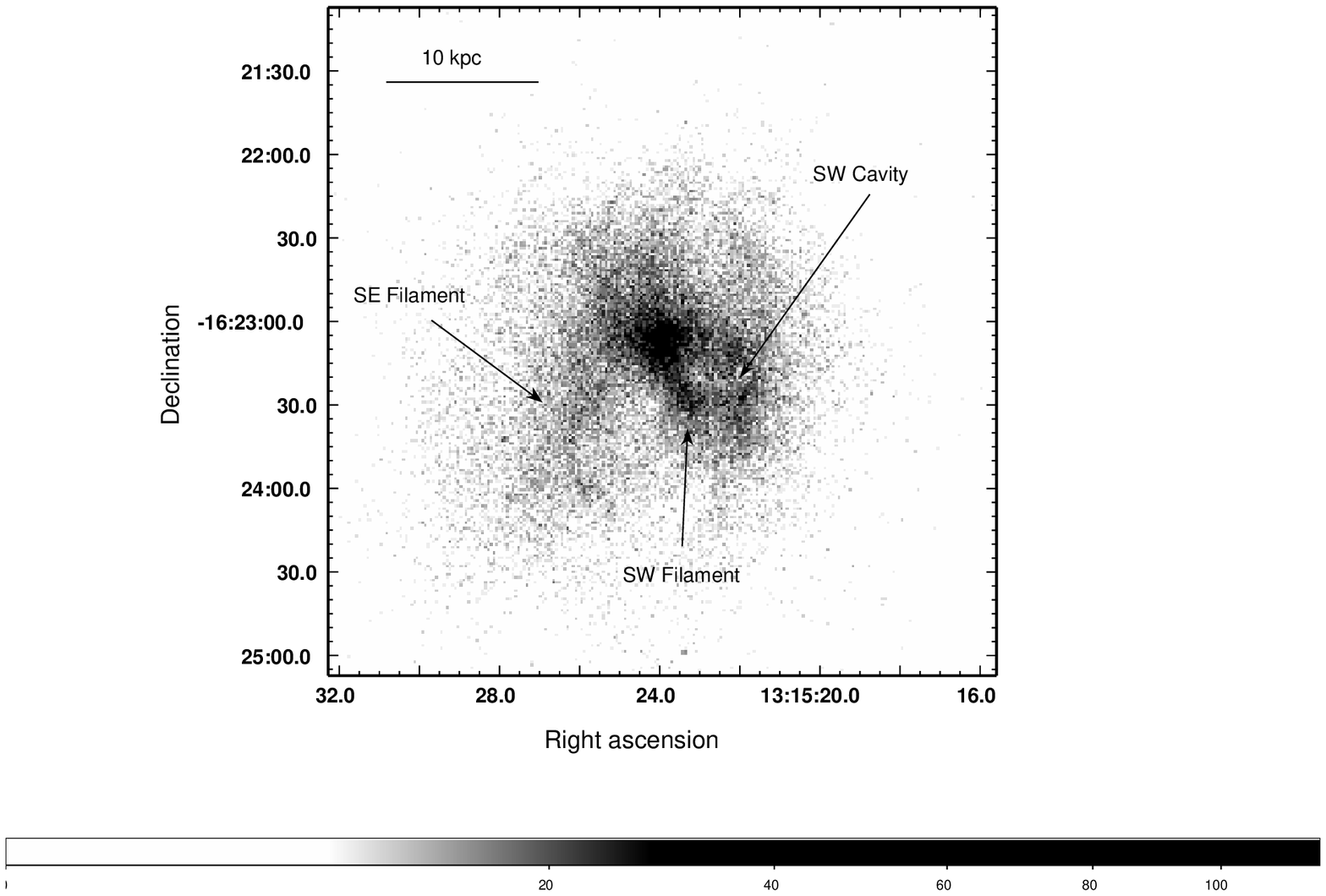}}
\caption{Raw 0.3-2.0~keV ACIS image of the central $4^{\prime}$ by $4^{\prime}$
(44~kpc by 44~kpc) region of NGC 5044.}
\end{inlinefigure}

\noindent
found evidence for multiphase gas within the central 30~kpc.
Gastaldello et al. (2008) noted the presence of two cavities, several 
cool filaments and a cold front approximately 30~kpc toward the south-east 
from NGC 5044 in the earlier ACIS observation.
The XMM-Newton observation of NGC 5044 showed that there is an additional cold front
approximately 50~kpc toward the north-west from NGC 5044 (Buote et al. 2003).
Analysis of the XMM-Newton RGS data for NGC 5044 by Tamura et al. (2003) found
that the emission from within the central 20~kpc is well represented by a two temperature
model with kT=0.7 and 1.1~keV. This range in temperatures is consistent
with the gas temperatures seen in projection based on the 
temperature profiles previously derived from ROSAT (David et al. 1995) 
and XMM-Newton EPIC (Buote et al. 2003) data.

This paper is organized in the following manner.  Details of the ACIS data analysis
are described in $\S 2$.  The complex X-ray morphology in the ACIS image is 
discussed in $\S 3$. In $\S 4$, we present temperature, pressure and entropy
maps of the central region of NGC 5044.  X-ray surface brightness profiles
are presented in $\S 5$ and gas temperature profiles are presented in $\S 6$.
Section 7 discusses the spectral analysis of the X-ray cavities and filaments.
In $\S 8$ we discuss the main implications of our results concerning the
AGN-cooling flow feedback mechanism and the possibility of sloshing induced
cold fronts in the center of the group.  A summary of our main results 
is given in $\S 9$.  Throughout this paper we use a luminosity distance 
of $D_L=38.8$~Mpc for NGC 5044, which corresponds to $1^{\prime\prime}=185$~pc.

\section{Chandra Data Analysis}

The NGC 5044 group of galaxies was observed by the Chandra X-ray Observatory 
on March 3, 2008 for a total of 82710 sec (OBSID 9399).  
The center of the group was positioned on the center of the back-side illuminated 
ACIS-S3 chip.  All data analysis was performed with CIAO 4.0.1 and 
CALDB 3.4.3.  This analysis includes the latest cti-corrected calibration products and
time dependent gain corrections for the 5 chips (S1,S2,S3,I2 and I3) that were turned on 
during the observation.  Since the X-ray emission from NGC 5044 completely fills the S3 chip, 
the back-side illuminated chip S1 was used to 

\begin{inlinefigure}
\center{\includegraphics*[width=1.00\linewidth,bb=53 196 483 589,clip]{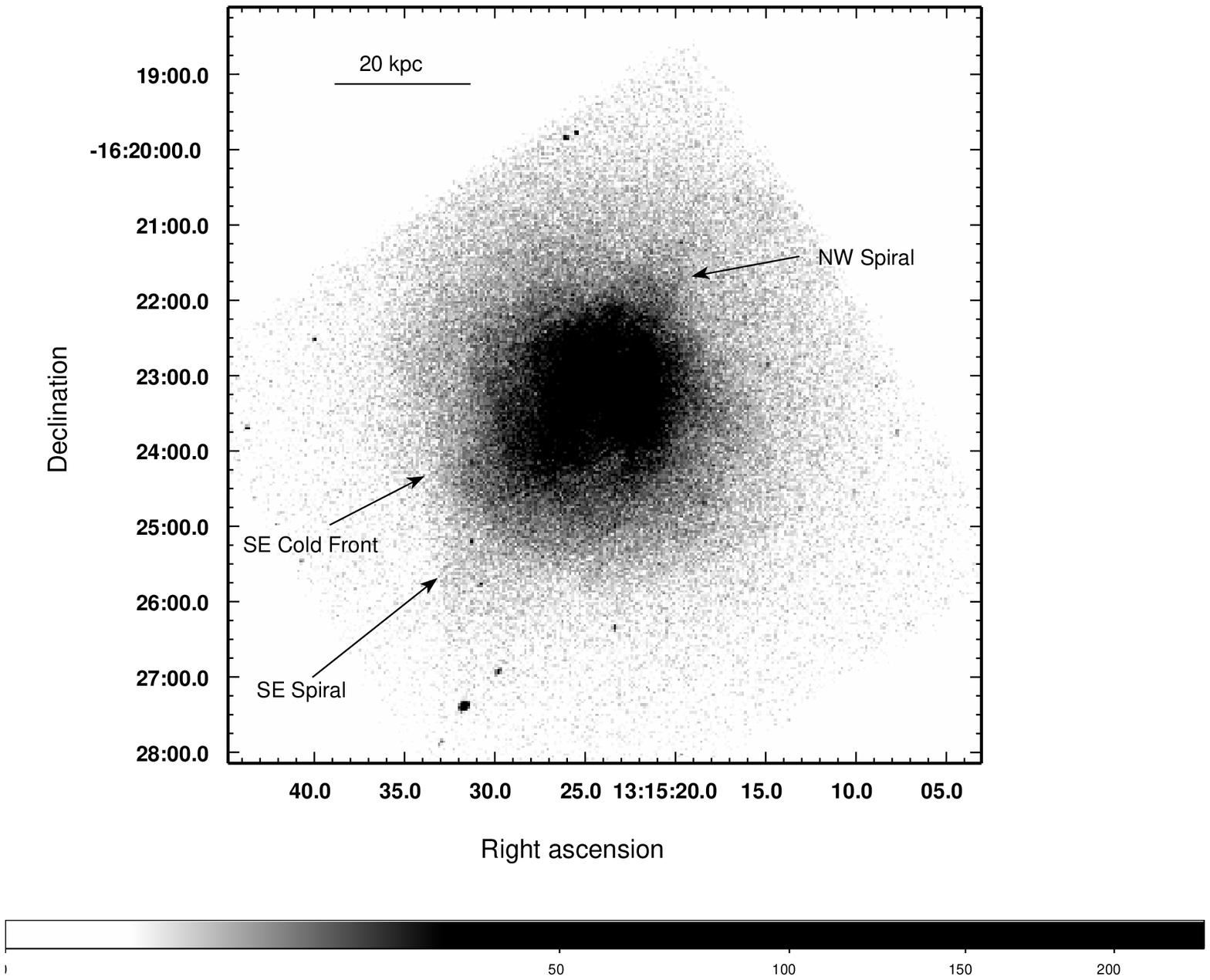}}
\caption{Raw 0.3-2.0~keV ACIS-S3 image of the central $10^{\prime}$ by $10^{\prime}$
(110~kpc by 110~kpc) region of NGC 5044.}
\end{inlinefigure}

\noindent
screen for background flares.  Light curves were made in 
the 2.5-7.0~keV and 9.0-12.0~keV energy bands for the diffuse emission on S1.  
No background flares exceeding a count rate threshold of 20\% above the quiescent background
count rate were detected during the observation.  NGC 5044 was observed near solar minimum, 
during which time background flares were significantly reduced.  Background images for 
the 5 chips that were turned on during the observation were extracted from the standard 
set of cti-corrected ACIS blank sky images in the Chandra CALDB.  The exposure time in 
each background image was adjusted to produce the same 9.0-12.0~keV count rate as that in 
the NGC 5044 observation.  The 9.0-12.0~keV count rates in the NGC 5044 observation were 
approximately 30\% higher than the rates in the standard background images, which is 
consistent with the increasing background rate during the few years leading up to solar 
minimum (Markevitch 2008).  The NGC 5044 observation was carried-out in very faint (VF) 
telemetry format and the VF background filtering was applied using the CIAO tool 
$acis\_process\_events$.  The same VF background screening 
was applied to the background data sets by only including events with "status=0".

\section{The Perturbed Inner Region of NGC 5044}

\subsection{Cavities, Filaments and Fronts}

The ACIS images in Figs. 1 and 2 show that the central region of NGC 5044 is highly 
perturbed with many cavities, filaments and edges.  Most of the cavities in NGC 5044 are 
fairly small, with diameters of only a few kpc, compared to the cavities found in 
rich clusters, which typically have diameters of tens of kpc (McNamara \& Nulsen 2007). 
Also, the inner cavities in NGC 5044 have a more isotropic distribution about the 
center of the group compared to the nearly bipolar distribution of the larger, 
more energetic, cavities seen in richer systems.  The central X-ray morphology of 
NGC 5044 is more similar to that of M87, which has many small cavities and 
filaments (Forman et al. 2007).  The largest filament identified in Fig. 1 
extends approximately 20~kpc toward the south-east.

\begin{inlinefigure}
\center{\includegraphics*[width=1.00\linewidth,bb=78 221 460 569,clip]{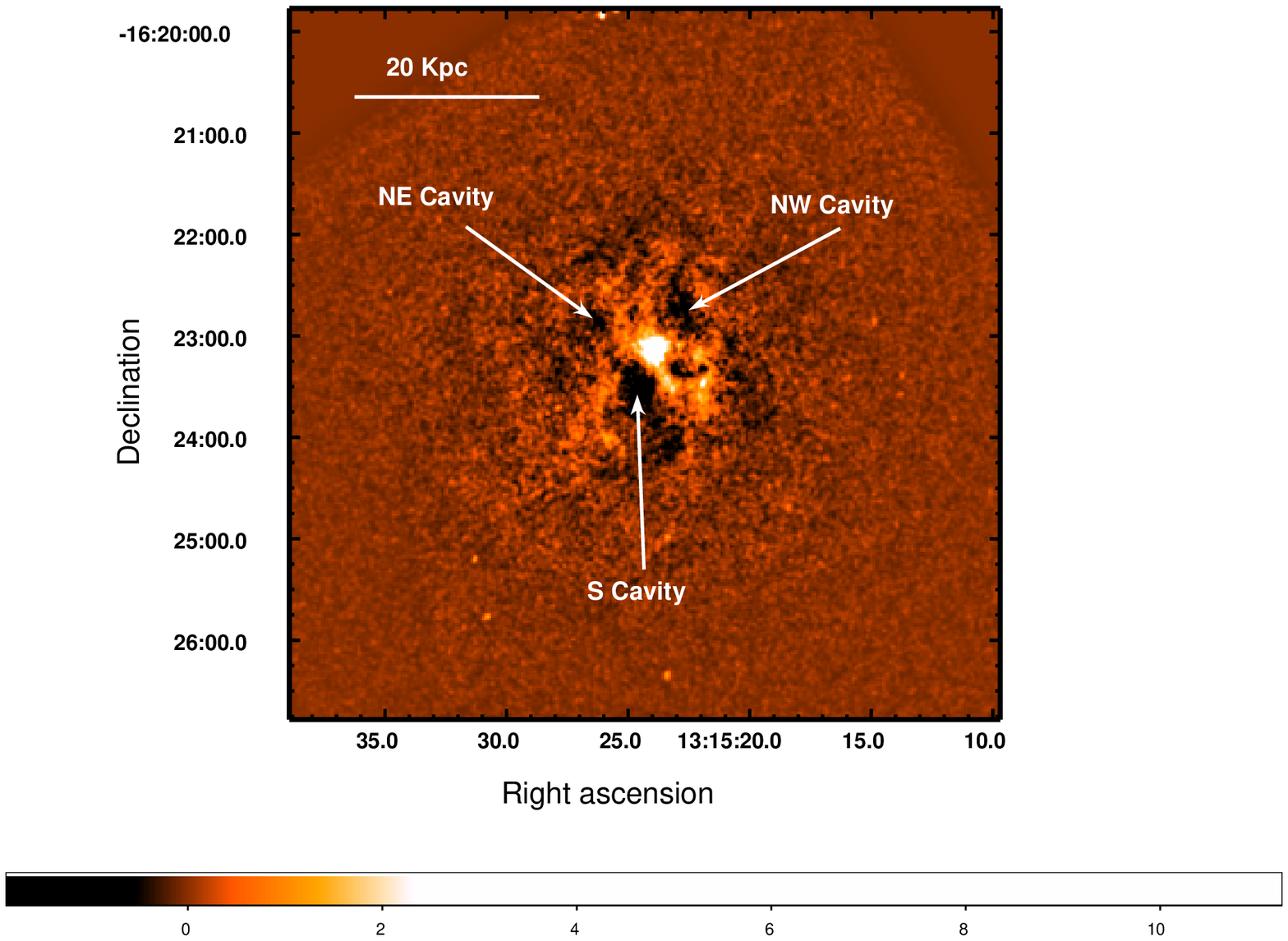}}
\caption{ACIS-S3 unsharp masked image of NGC 5044 in the 0.3-2.0~keV band.}
\end{inlinefigure}

Beyond the central cavities and filaments, there are two edges toward the south-east and one
toward the north-west (see Fig. 2).  The spectral analysis presented 
below (see $\S 6$) shows that 
all of these edges are cold fronts and not shocks 
(i.e., the gas interior to the fronts is cooler than the surrounding gas).  
The innermost cold front located 30~kpc toward the south-east
was noted by Gastaldello et al. (2008) based on the earlier Chandra observation.  
Another cold front is also visible toward the south-east in Fig 2 with more of 
a spiral pattern and extends 
approximately 50~kpc toward the south.  A corresponding spiral shaped cold front 
is also seen in Fig. 2 toward the north-west. This spiral feature is more obvious in the 
surface brightness analysis presented below. Cold fronts are commonly found 
near the central dominant galaxy in clusters and may be due to merger induced sloshing 
of the central galaxy (Markevitch \& Vikhlinin 2007; Ascasibar \& Markevitch 2006).  
However, the two-arm spiral pattern in NGC 5044 is more similar to the AGN inflated 
"hour glass" shape feature observed in NGC 4636 (Jones et al. 2002; O'Sullivan, 
Vrtilek \& Kempner 2005; Baldi et al. 2009). The origin of the cold fronts in NGC 5044 
will be discussed in more detail below.  The full complexity of the X-ray morphology 
of NGC 5044 is best revealed in the unsharp masked image shown in Fig. 3. This image 
shows the full extent of the X-ray cavities and the enhancements in the surface 
brightness around the cavities.  There are also several larger and weaker depressions in 
the unsharp masked image beyond the cavities identified in Fig. 3.
It is obvious from Fig. 3 that the cavities occupy 
a significant fraction of the total volume within the central 10~kpc.

\subsection{Radio Morphology}

NGC 5044 was recently observed by the Giant Metrewave Radio Telescope (GMRT)
at frequencies of 235, 325 and 610~MHz as part of a sample of groups (Giacintucci et al. 2009a).  
Previous higher frequency observations of NGC 5044 only detected a compact radio source 
with a flat spectral index consistent with a core-dominated source (Sparks et al. 1984).  
The GMRT observation at 610~MHz shows the presence of extended emission toward the 
south-east with a total flux of 8.5~mJy (see Fig. 4). 

\begin{inlinefigure}
\center{\includegraphics*[width=1.00\linewidth,bb=94 233 455 564,clip]{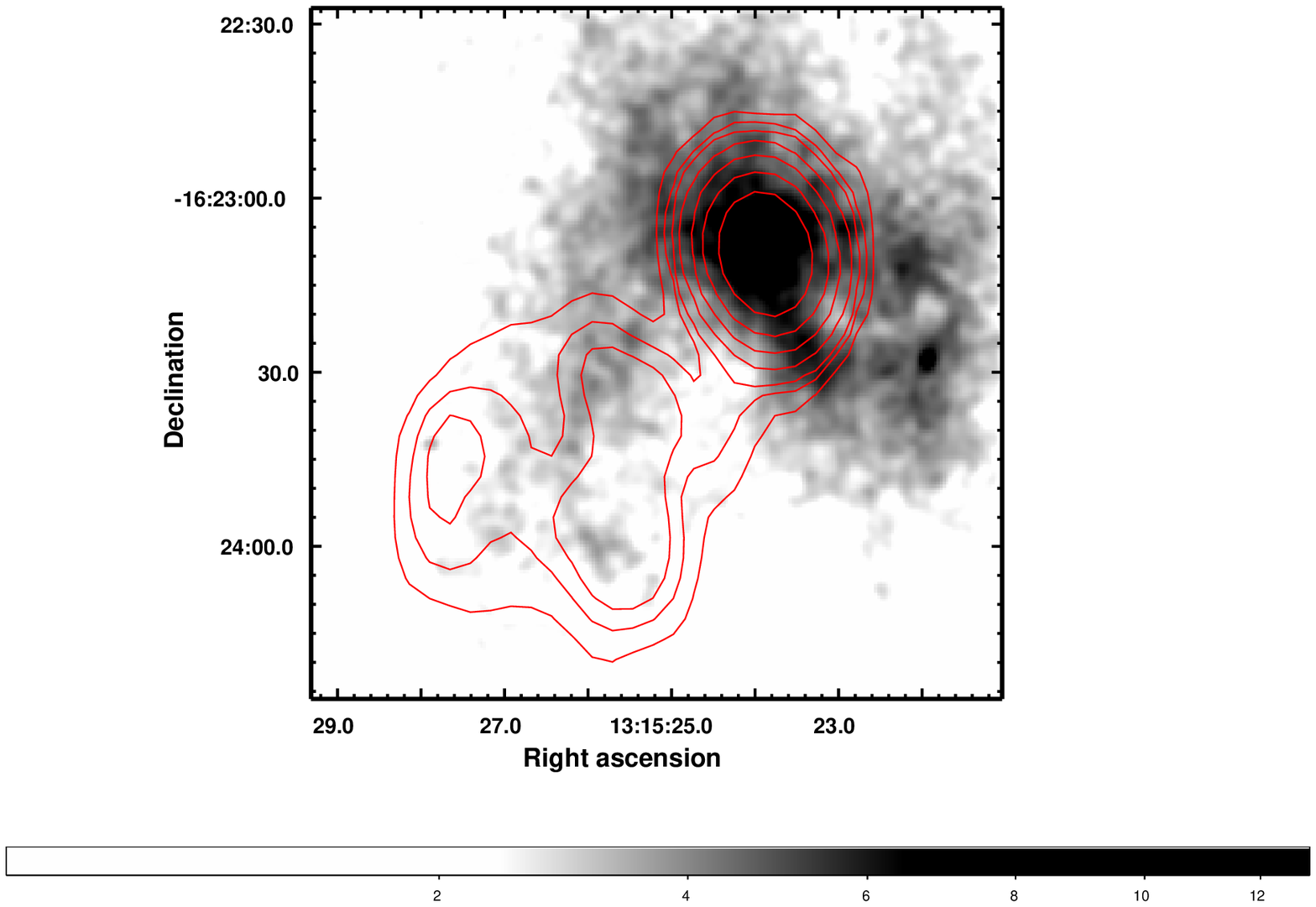}}
\caption{GMRT 610~MHz contours overlayed on the raw 0.3-2.0~keV ACIS image. The beam
size is 18$^{\prime\prime}$ by 13$^{\prime\prime}$ and the lowest radio contour is
shown at $3 \sigma$ = 0.2mJy/b.}
\end{inlinefigure}

\noindent
The two off-set peaks in the 610~Mhz radio image visible in Fig. 4 suggest
that the 610~MHz radio emission arises from within a torus shaped region with the
axis of the torus in the plane of the sky.
This figure also shows that the cool 
gas in the SE filament appears to be threading the center of the torus.  
This behavior is very similar to that 
seen in numerical simulations of buoyantly 
rising AGN inflated bubbles which develop into torus-like structures with cool gas 
from the center of the cluster being dredged up in their wakes 
(Churazov et al. 2001; Gardini 2007; Revaz et al. 2008).
This phenomenon is observed in the eastern arm of M87 (Forman et al. 2007)
and in the dredging up of H$~\alpha$ filaments behind AGN inflated bubbles in 
the Perseus cluster (Hatch et al. 2006).  

The radio emission at 235~MHz is much more extended than the emission 
at 610~MHz and there is little overlap between the two frequencies (see Fig. 5).
The lack of any detected emission at 610~MHz in the same region as the 235~MHz
emission indicates that the radio spectrum must be very steep with a spectral index 
of $\alpha \gax 1.6$. The absence of any 610~MHz emission in the 235~MHz radio lobes 
implies that the radio spectrum of the lobes must be fairly flat ($\alpha \lax 1.6$).
There are two separate extended components in the 235~MHz observation.
One of the components originates at 
the center of NGC 5044, passes through the southern cavity and then bends toward 
the west just behind the SE cold front (see Fig. 6).  The bending is probably due 
to the streaming of gas behind the SE cold front.  Beyond the SE cold front, the radio
emission sharply bends toward the south, possibly due to the effects of buoyancy.
While the cavities in NGC 5044 have a nearly isotropic distribution,
Fig. 6 shows that all of the extended radio emission lies to the 
south and south-east.  Fig 6. also shows that no radio emission is detected
in most of the X-ray cavities.

The GMRT observation at 235~MHz shows that there is a second detached radio
lobe toward the south-east (see Figs. 5 and 6).
The north-western edge of the this radio lobe coincides with the SE cold front, suggesting
that the relativistic material in the lobe was produced by an earlier radio outburst

\begin{inlinefigure}
\center{\includegraphics*[width=1.00\linewidth,bb=142 283 408 516,clip]{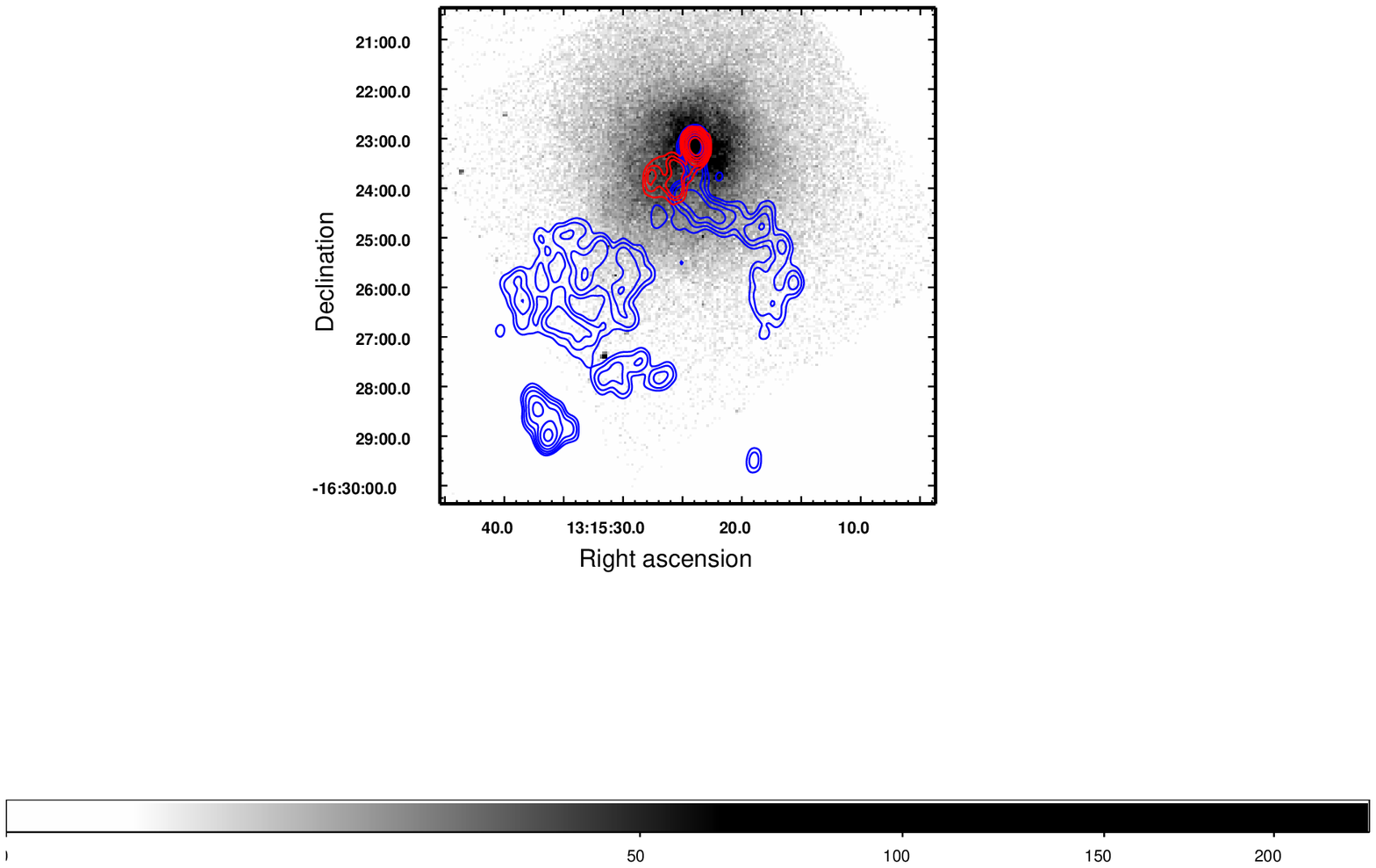}}
\caption{GMRT 610~MHz contours (red) and 235~MHz contours (blue) overlayed on the
raw 0.3-2.0~keV ACIS image. The 610~MHz contours are the same as those shown in Fig. 4.
For the 235~MHz data, the beam size is 22$^{\prime\prime}$ by 16$^{\prime\prime}$
and the lowest contour is shown at $3\sigma = 0.75$~mJy~b$^{-1}$. The source in the
bottom left corner is a background radio galaxy.}
\end{inlinefigure}

\noindent
from NGC 5044 
and that it is currently be compressed by the motion of NGC 5044 toward the south-east.
The GMRT data thus appears to reveal 2 or possibly 3 separate radio outbursts. The
youngest outburst can be
identified with the 610~MHz emission and the 
oldest outburst with the detached radio lobe.
The GMRT observation of NGC 5044 will be discussed in more detail in
Giacintucci et al. (2009b).

\subsection{Point Sources}

The central AGN and other hard X-ray point sources are easily seen 
in the raw 2.0-7.0~keV ACIS image (see Fig. 7).  Using the CIAO tool {\it wavdetect}, 
12 point sources, in addition to the central AGN, are  detected within the 
central 4$^{\prime}$ by 4$^{\prime}$ (44 by 44~kpc) region with an unabsorbed 
2.0-7.0~keV flux limit of $1.26 \times 10^{-15}$~erg~cm$^{-2}$~s$^{-1}$ (assuming 
a power-law spectrum with an index of $\Gamma=1.4$).  The hard X-ray band log N-log S 
relation for background point sources given by Moretti et al (2003) predicts $7 \pm 3$ 
point sources above this flux limit within this region, so there is not a significant 
excess of point sources over this region.  However, within the central 15~kpc, there 
are 8 detected point sources in addition to the central AGN and the predicted background 
number is $2.3 \pm 1$.  Thus, some of these point sources are probably low-mass X-ray 
binaries associated with NGC 5044.
Assuming these 8 point sources are at the distance of NGC 5044, gives 2-7~keV
luminosities between $3.3 - 15.3 \times 10^{38}$~erg~s$^{-1}$, which are typical
luminosities for the most luminous LMXBs in early-type galaxies (e.g., Sarazin, 
Irwin \& Bregman 2001; Kraft et al. 2001; Kim \& Fabbiano 2004; David et al. 2006; 
Brassington et al. 2008).

\section{Temperature, Pressure and Entropy Maps}

Most of the X-ray emission from 1~keV groups and early-type galaxies
arises from L shell transitions from Fe with ionization states from 
Fe XIX (Ne-like) through Fe XXIV (He-like).  With CCD spectral resolution,
these L-shell lines are blended into a single feature between 
approximately 0.7 and 1.2~keV.  When fitting the spectra of gas with temperatures
of approximately 1~keV, the temperature is primarily determined

\begin{inlinefigure}
\center{\includegraphics*[width=1.00\linewidth,bb=90 228 455 562,clip]{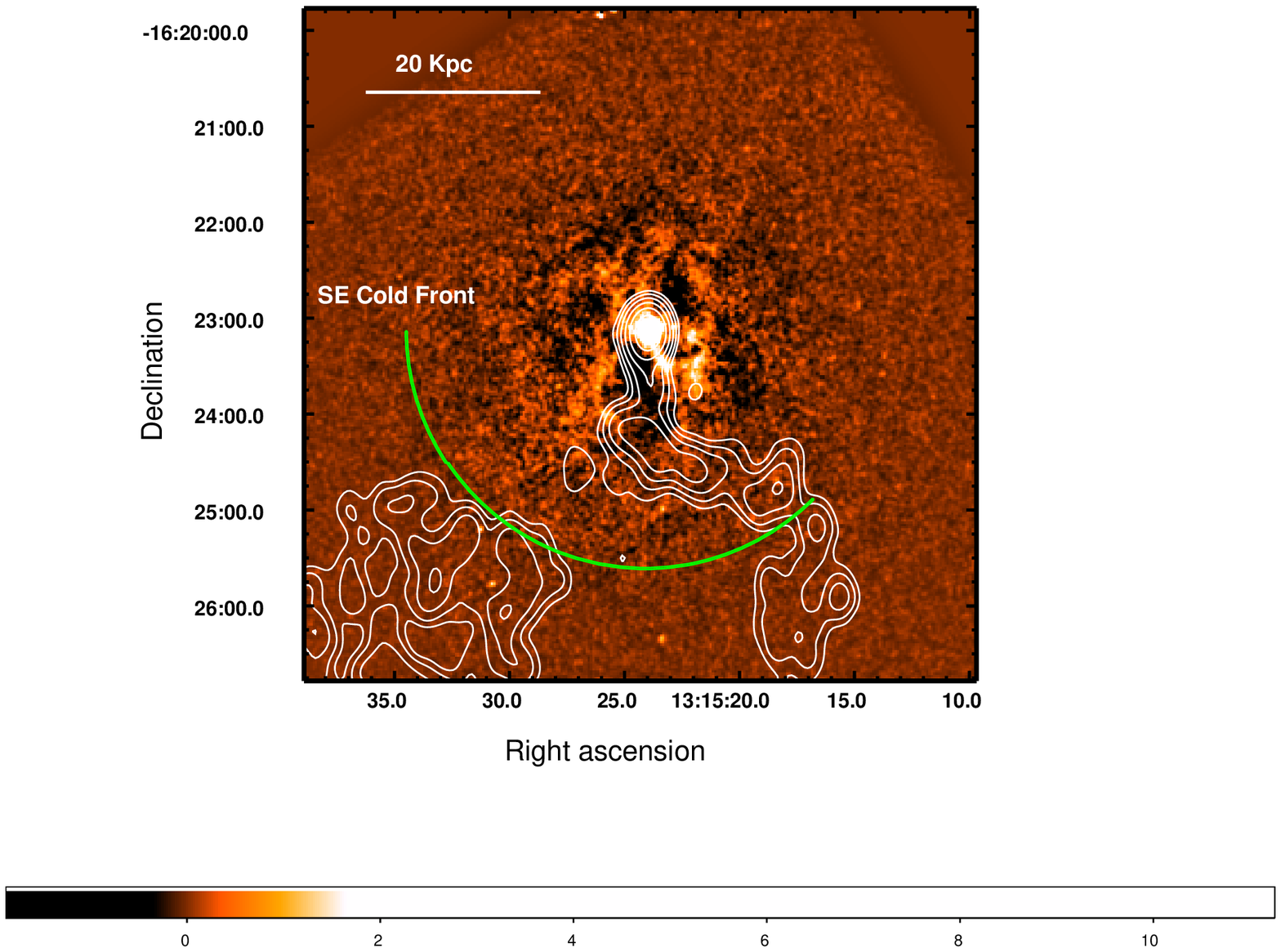}}
\caption{GMRT 235~MHz contours overlayed on the unsharp masked image.  The contours are
the same as shown in Fig. 5. Also shown is the location of the SE cold front.}
\end{inlinefigure}

\noindent
by the energy centroid of the blended Fe-L lines. 
This results from the shift in the dominant 
ionization state of Fe from Fe XIX in 0.5~keV gas to Fe XXIV in 1.2~keV gas.  
Since Li-like Fe is the highest ionization state of Fe that can produce L shell
lines, the mean photon energy of the blended L shell lines is independent of
temperature above 1.2~keV. 
To determine the correlation between mean photon energy in the 0.7-1.2~keV
energy band and gas temperature, we generated a series of XSPEC simulated spectra 
based on an absorbed vapec 
model with temperatures between 0.5 and 1.5~keV.  
These simulations assume galactic absorption ($N_{gal}=4.94 \times 10^{20}$~cm$^{-2}$) 
and the redshift of NGC 5044.  The emission between 0.7 and 1.2~keV also contains 
some emission from O, Ne and Mg, however, assuming solar abundances for all 
elements, 93\% of the emission in this energy band arises from Fe in 1~keV gas, 
so the mean photon energy is essentially independent of the assumed abundance ratios. 
Fig. 8 shows that there is a linear correlation
($kT=-4.61+5.88 <E>$) between temperature and mean photon energy
up to temperatures of approximately 1.2~keV.  

Using the linear correlation shown in Fig. 8
between mean photon energy and temperature between 0.5 and 1.2~keV (which is
approximately the maximum observed temperature in NGC 5044), we have 
generated a high resolution temperature map of NGC 5044 (see Fig. 9).  The
advantage of this method is that
far fewer counts are required to compute
the mean photon energy than that required for a full spectral analysis. Also, the 
derived temperature is independent of the Fe abundance and abundance ratios.
The temperature map shown in Fig. 9 was derived by first applying a $3 \sigma$ adaptive
smoothing to the raw image in the 0.7-1.2~keV band.  The same kernel was then used
to adaptively smooth an image of the summed photon energy in the 0.7-1.2~keV band.
Dividing these images produces an image of the mean photon energy and the temperature
can then be computed from the linear relation shown in Fig. 8. 
The primary source
of uncertainty in the estimated temperatures is the discrepency between the linear
function used in Fig. 8 and the mean photon energies computed from the XSPEC simulations
and corresponds to about 2-3\%.

\begin{inlinefigure}
\center{\includegraphics*[width=1.00\linewidth,bb=85 215 460 555,clip]{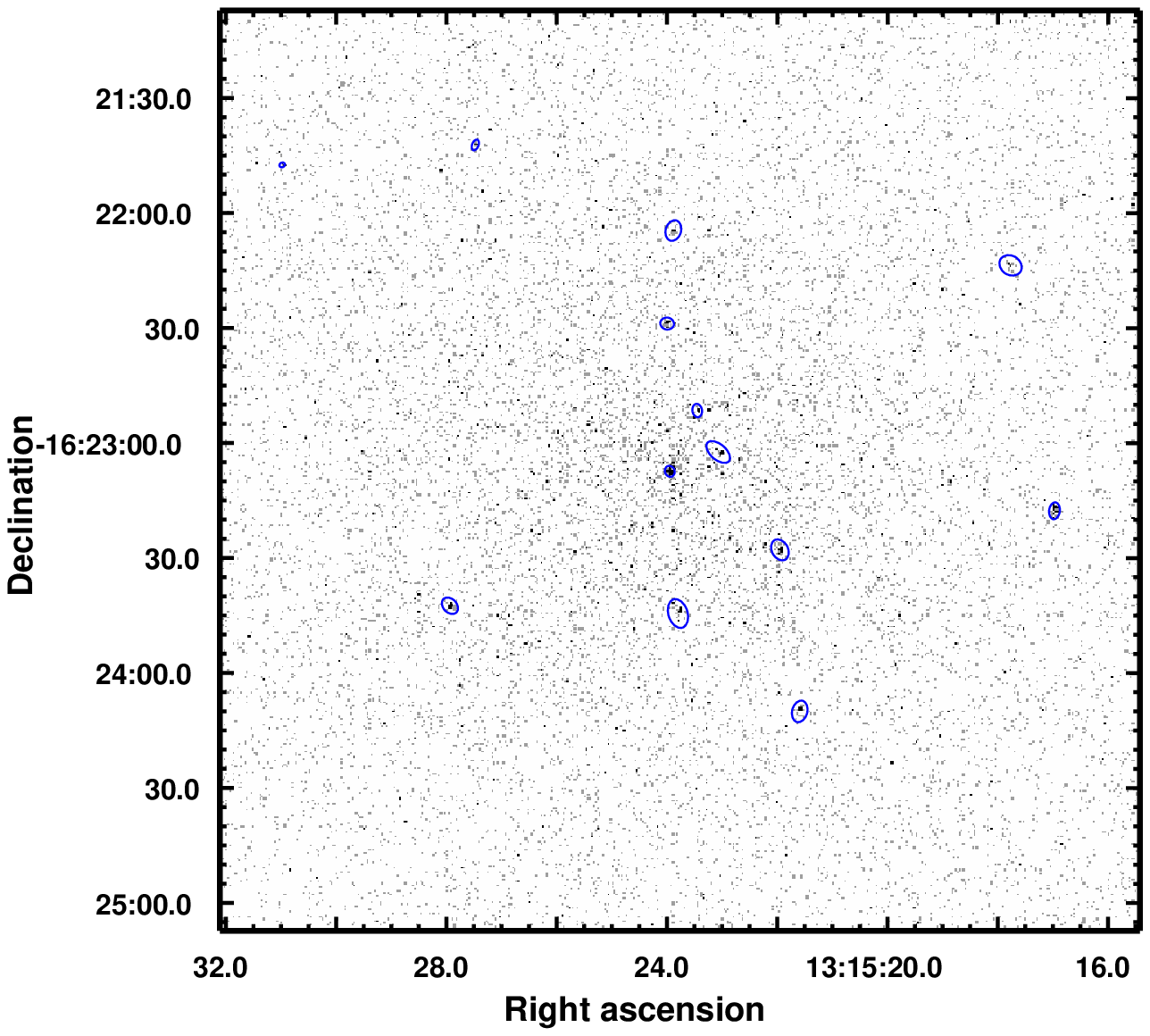}}
\caption{Raw 2.0-7.0~keV ACIS image of the central $4^{\prime}$ by $4^{\prime}$
region of NGC 5044. All detected point sources are identified with elliptical contours.}
\end{inlinefigure}

\begin{inlinefigure}
\center{\includegraphics*[width=1.00\linewidth,bb=21 148 587 695,clip]{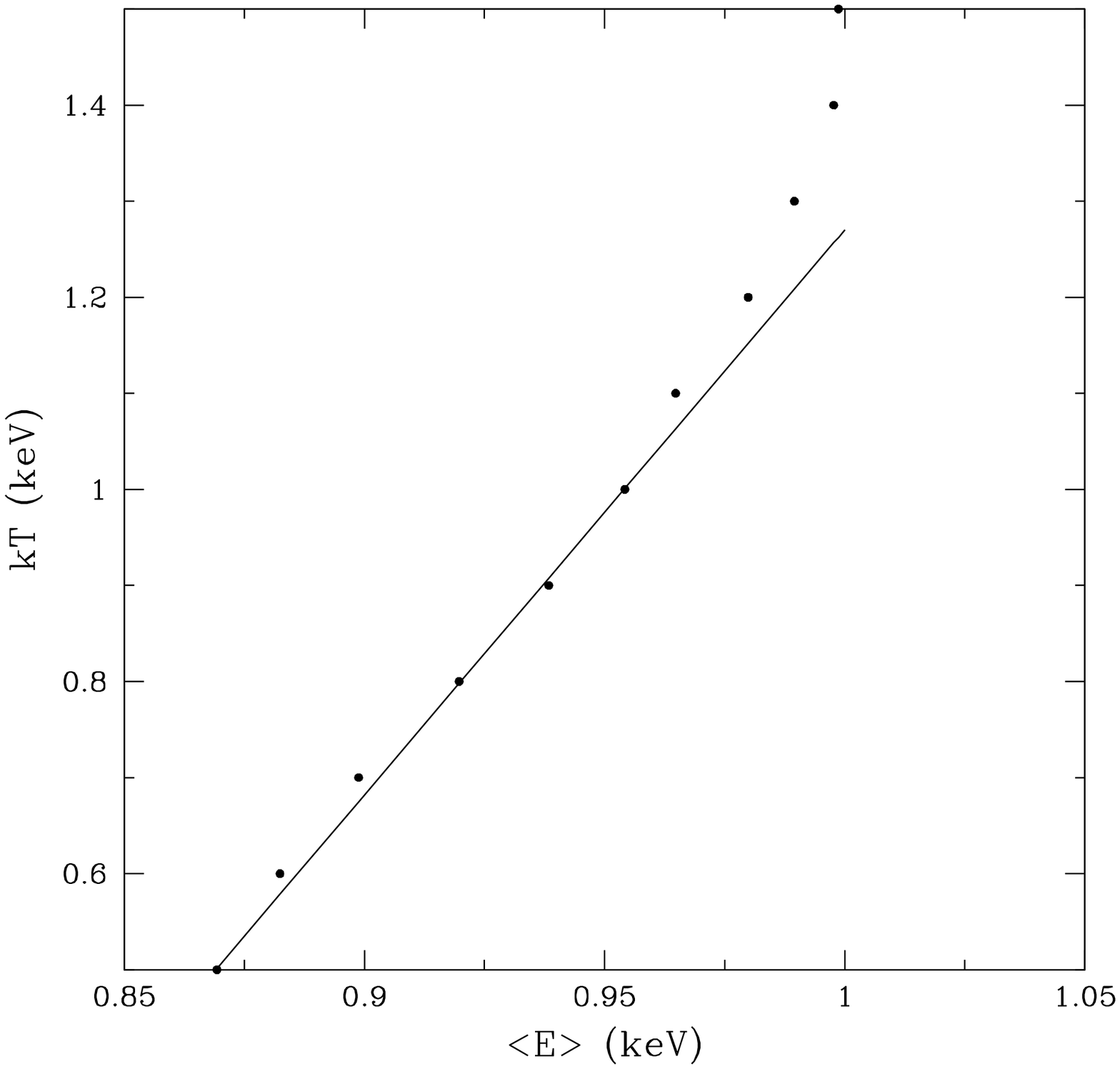}}
\caption{Correlation between gas temperature and mean photon energy in the
0.7-1.2~keV energy band as derived from XSPEC simulations.}
\end{inlinefigure}

Fig 9. shows that there is a great deal of structure in the gas temperature within 
the central 50~kpc of NGC 5044.  There is a clear extension of the coolest gas toward
the south-east in the direction of the SE cold front (shown by the white arc in Fig. 9)
and the SE spiral.  A hardness ratio image of the ROSAT observation of NGC 5044 also showed
that the X-ray softest emission in NGC 5044 was contained in a "tear-drop" shaped
feature extending toward the south-east (David et al. 1995). The SE cold front 
clearly bounds most of the cool central gas, however, the cool gas in the SE
spiral can be detected out to approximately 50~kpc from the AGN, which is beyond 
any detected radio emission.  All of the bright filaments visible in Figs. 1 
and 2 have cooler gas compared to the surrounding regions.  Higher temperatures are 
observed in the cavities, probably due 

\begin{inlinefigure}
\center{\includegraphics*[width=1.00\linewidth,bb=100 192 441 559,clip]{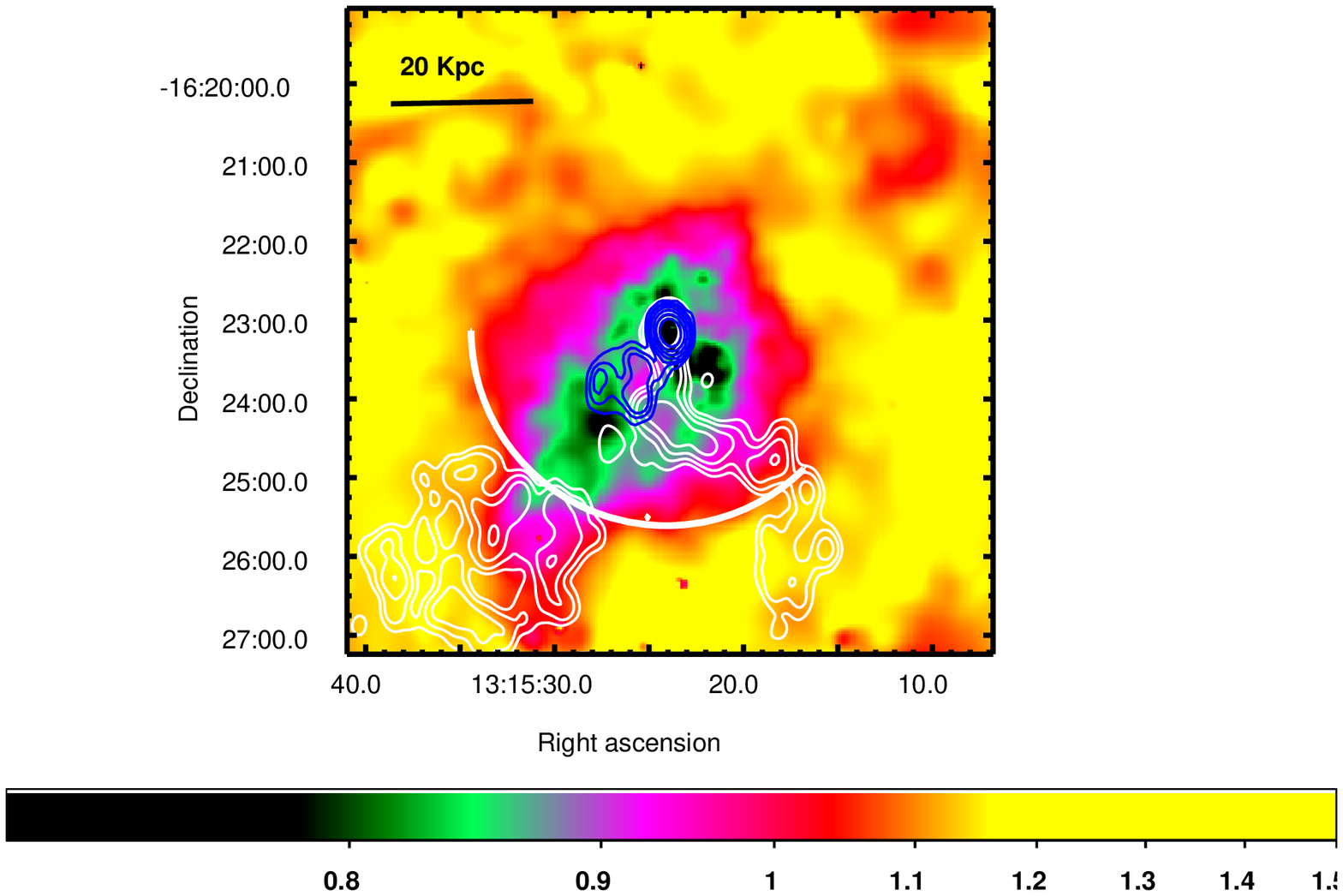}}
\caption{Temperature map of the central region of NGC504.  The color coded
gas temperatures are shown on the color bar.  Also shown are the GMRT contours at
610~MHz (blue) and 235~MHz (white) and the location of SE cold front.}
\end{inlinefigure}

\noindent
to foreground and background emission from hotter gas. 
The fact that the colors in the temperature map at the locations of the cavities are red 
(0.9-1.0~keV) and not yellow (1.1-1.2~keV), indicates that these cavities are not 
highly elongated along the line-of-sight.  The 235~MHz and 610~MHz radio contours 
show that the AGN is located at the north-east end of the cool SW filament.  

The correspondence between gas temperature and enhancement in the X-ray surface 
brightness is shown in Fig. 10. This 
figure plots the gas temperature, as determined from 
the temperature map, and the value in the unsharp masked image for every point
in the two images.  Positive values in the unsharp masked image correspond to 
enhancements in the surface brightness 
and negative values correspond to depressions.  
Below temperatures 
of 0.9~keV, Fig. 10 shows that there is a general trend for 
enhanced regions to contain cooler gas and for depressions to contain 
hotter gas, suggesting that these regions are in rough 
pressure equilibrium. 
One of the more surprising results in Fig. 10 is the lack of significant 
enhancements or depressions in gas 
hotter than 0.9~keV.  Based on the temperature
profiles presented below, this corresponds to a radial distance of approximately
20-30~kpc.  We also produced the same plot derived from unsharp masked images 
generated from a range of smoothing lengths and found the same result.
The trend in Fig. 10 shows 
that there are no cavities or filaments 
of the same spatial scale or amplitude as those seen in the central region beyond the 
SE cold front.  This suggests that AGN inflated, buoyantly rising 
cavities with sizes similar to those currently seen in the center 
of NGC 5044 are disrupted, or prevented from buoyantly rising
beyond the central 30~kpc.

A pressure map of the central region of NGC 5044 was generated from the product
of the temperature map shown in Fig. 9 and a density map.  The density map was
derived from the 

\begin{inlinefigure}
\center{\includegraphics*[width=1.00\linewidth,bb=20 145 572 703,clip]{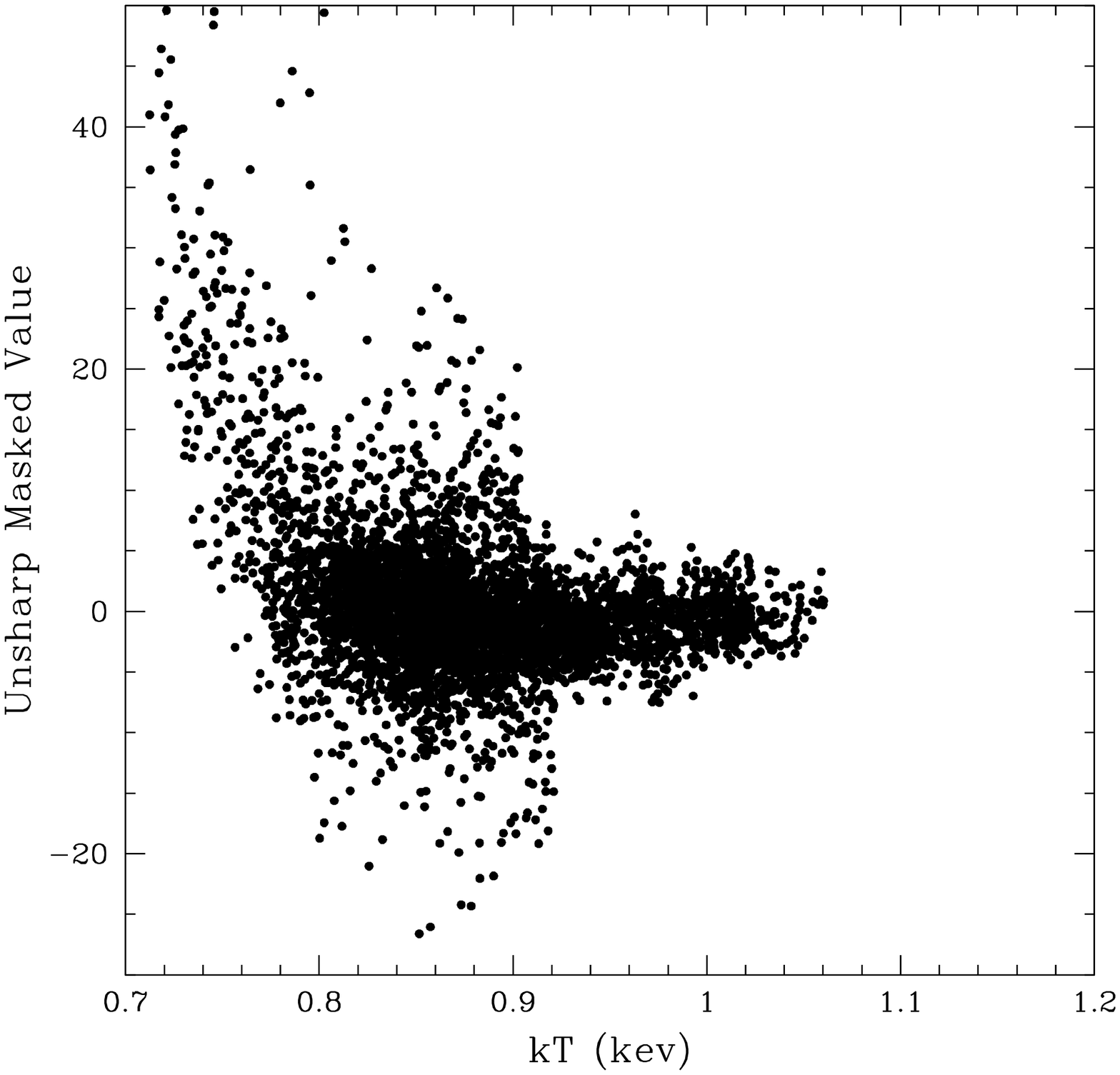}}
\caption{Comparison between the gas temperature (determined from the temperature
map in Fig. 9) and the enhancement in the X-ray surface brightness based on the value
in the unsharp masked image shown in Fig. 3.}
\end{inlinefigure}

\begin{inlinefigure}
\center{\includegraphics*[width=1.00\linewidth,bb=156 287 404 510,clip]{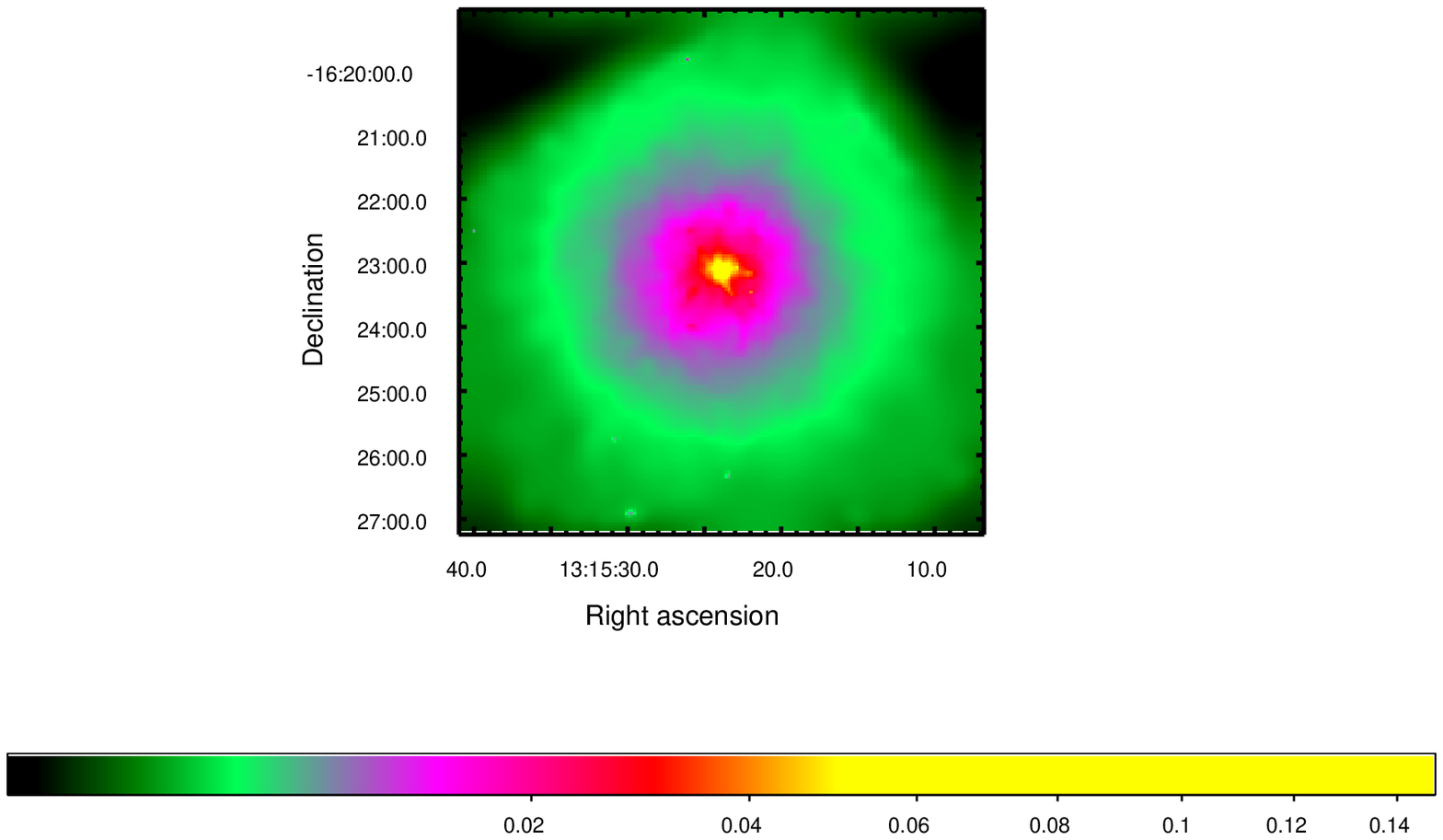}}
\caption{Gas pressure map of the central region of NGC 5044.}
\end{inlinefigure}

\noindent
same $3 \sigma$ adaptively smoothed image used to calculate the
temperature map assuming $n_e \propto (\Sigma/r)^{1/2}$, where $\Sigma$ is the surface
brightness in a given region and $r$ is the distance between the given region and the center 
of the group.  The electron density was normalized at a radial distance of 
100~kpc based on the results of a deprojected spectroscopic analysis.  The deprojected
spectral analysis will be presented in a subsequent paper (David et al. 2009).
Even though there is significant structure in the ACIS image and temperature map,
the resulting pressure map shown in Fig. 11 is nearly azimuthally symmetric with only 
a few weak features.  To enhance any pressure residuals, we computed an
azimuthally averaged pressure map based on 

\begin{inlinefigure}
\center{\includegraphics*[width=1.00\linewidth,bb=157 251 398 513,clip]{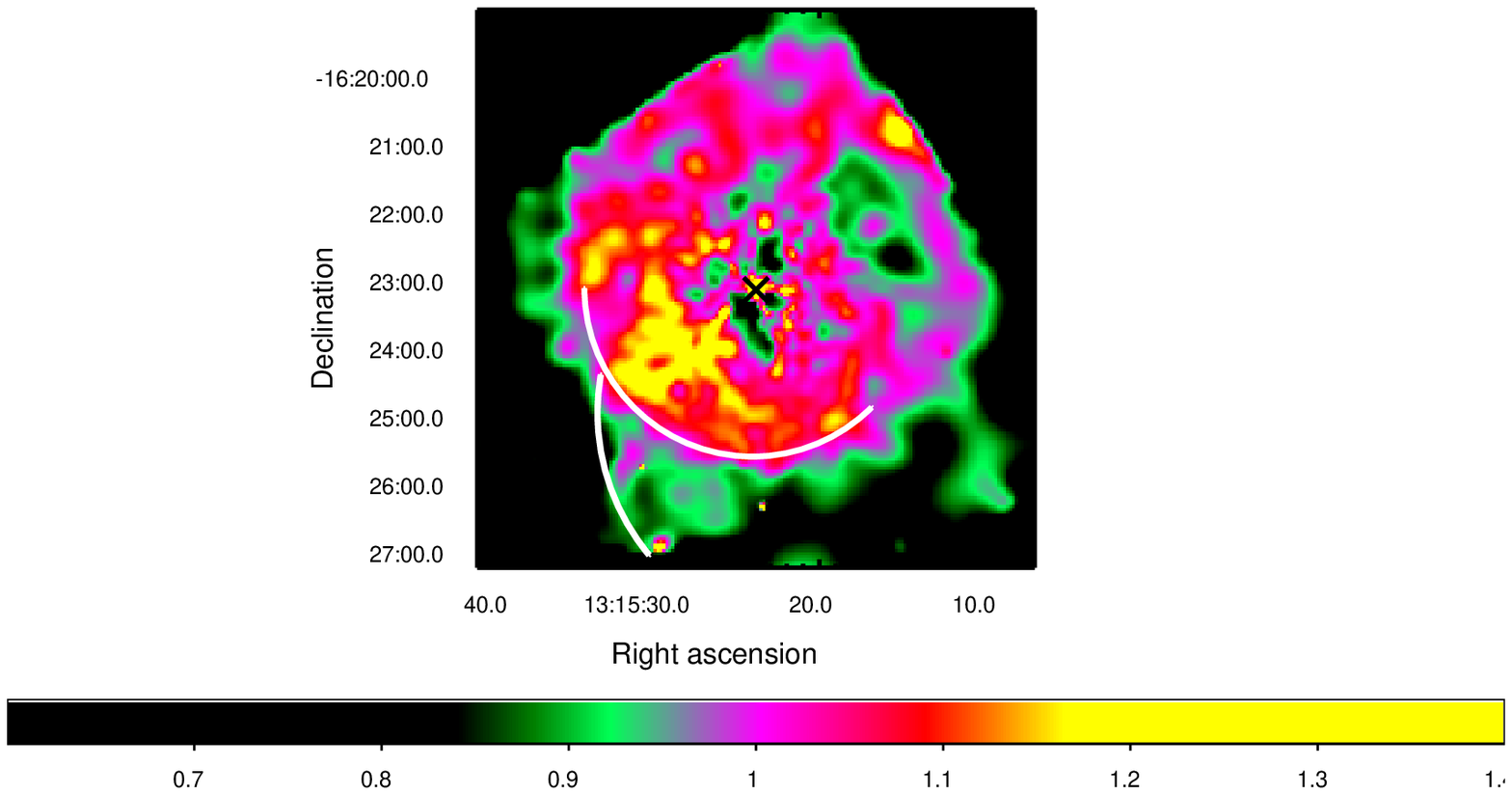}}
\caption{Pressure residual map of the central region of NGC 5044. The positions of
the SE cold front and SE spiral are shown as white arcs and the location of the
central AGN is indicated with a "X". The relative magnitude of the pressure residuals are
shown on the color bar.}
\end{inlinefigure}

\noindent
the azimuthally averaged pressure 
profile derived from the pressure map.  A pressure residual map (see Fig. 12) was then 
obtained by dividing the pressure map with the azimuthally averaged pressure map.
Fig. 12 shows that the pressure is approximately 20\% higher toward the south-east and 
20\% lower toward the north-west relative to the azimuthally averaged values.
The outer edge of the high pressure region toward the south-east coincides with the SE
cold front shown as a white arc in the Fig. 12, providing further evidence 
that this cold front is due to motion of NGC 5044 relative to the group gas.

An entropy map, where $S = kT/n_e^{2/3}$, was computed directly 
from the temperature 
and density maps (see Fig. 13).  The overall shape of the entropy map is similar
to the temperature map with the lowest entropy gas more extended in the direction
of the SE cold front and spiral.  Fig.13 shows that the entropy gradient
is the steepest near the SE cold front.

\section{Surface Brightness Profiles}

\subsection{Azimuthally Averaged Profile}

Emission from the NGC 5044 group was detected on all 5 CCDS that were
turned on during the ACIS observation and data is available out to a radius of 
approximately $20^{\prime}$ (220~kpc).  Fig. 14 shows the background-subtracted
and exposure corrected surface brightness profile in the 0.3-2.0~keV energy
band including data from all 5 CCDs.  The background for each CCD 
was determined 
using the method described in $\S 2$.  Separate exposure maps were generated for 
each CCD at an energy 
corresponding to the mean source photon energy
in the 0.3-2.0~keV 
energy band. The central point source is evident in the innermost
bin of the surface brightness profile shown in Fig. 14, as are the inner cavities
and a central power-law region extending out to approximately 3~kpc
with a slope of $\sim 0.3$. Even though

\begin{inlinefigure}
\center{\includegraphics*[width=1.00\linewidth,bb=157 290 401 510,clip]{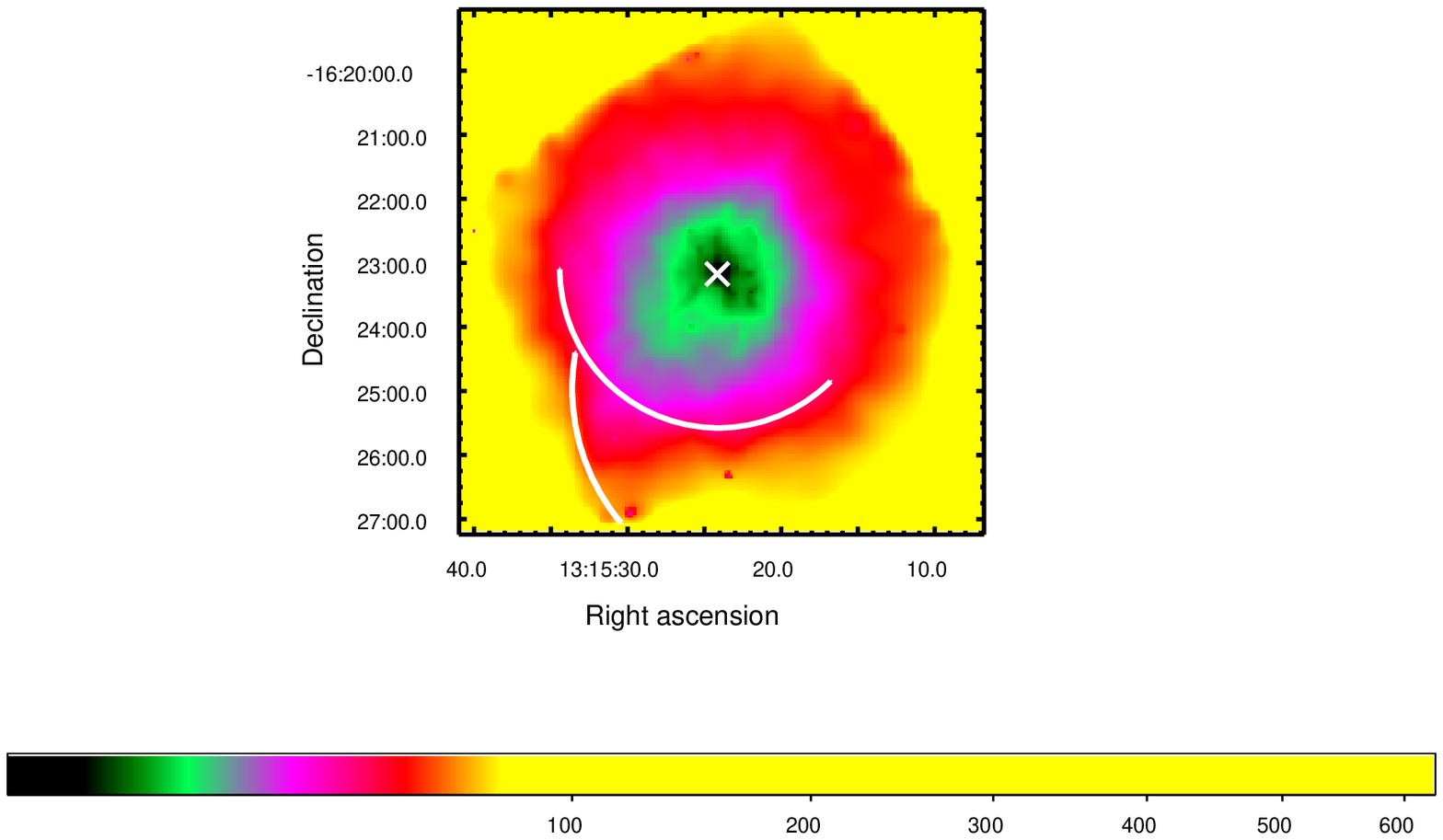}}
\caption{Entropy map of the central region of NGC 5044. The positions of
the SE cold front and SE spiral are shown as black arcs and the location of the
central AGN is indicated with a "X". The innermost contour is drawn at an entropy of
10~keV~cm$^2$ and the contour levels increases in increments of 10~keV~cm$^2$.}
\end{inlinefigure}

\begin{inlinefigure}
\center{\includegraphics*[width=1.00\linewidth,bb=20 145 569 695,clip]{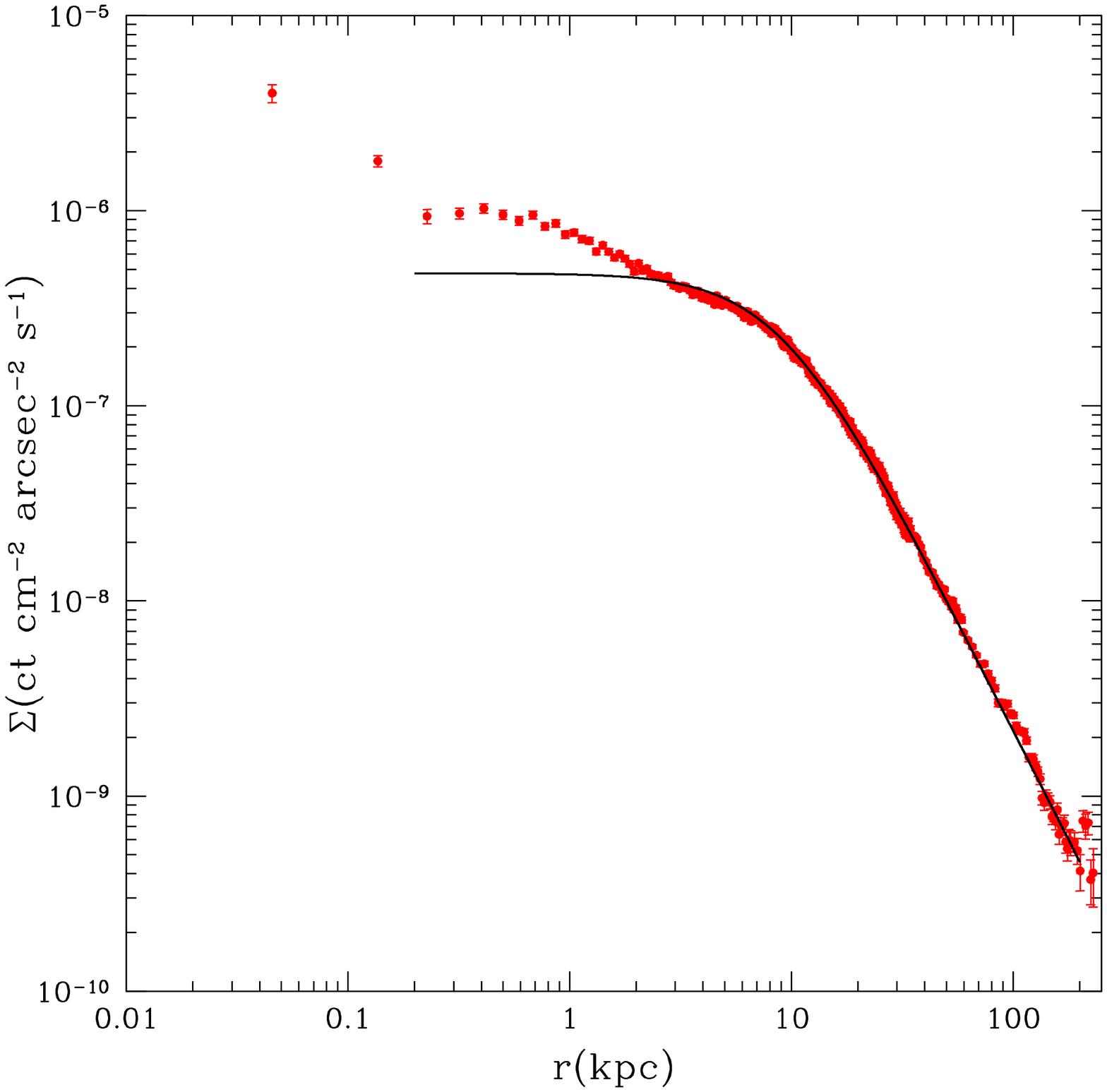}}
\caption{Azimuthally averaged, background-subtracted and exposure corrected 0.3-2.0~keV surface
brightness profile of NGC 5044. The best fit $\beta$ model is shown as a solid line.}
\end{inlinefigure}

\noindent
there is significant substructure in the central
region of NGC 5044, beyond 3~kpc, the azimuthally averaged surface brightness
profile is remarkably smooth.  We therefore fitted the 0.3-2.0~keV surface brightness 
profile beyond 3~kpc to the standard $\beta$ model:

$$\Sigma(r) = \Sigma_0 \left[ 1 + \left( { {r} \over {r_0}} \right)^2 \right]^{-3 \beta + 0.5}$$

\noindent 
and obtained $r_0=9.1 \pm 0.1$~kpc and $\beta=0.541 \pm 0.001$ (1 $\sigma$ errors).
The outer slope of the best-fit $\beta$ model shown in

\begin{inlinefigure}
\center{\includegraphics*[width=1.00\linewidth,bb=20 145 569 695,clip]{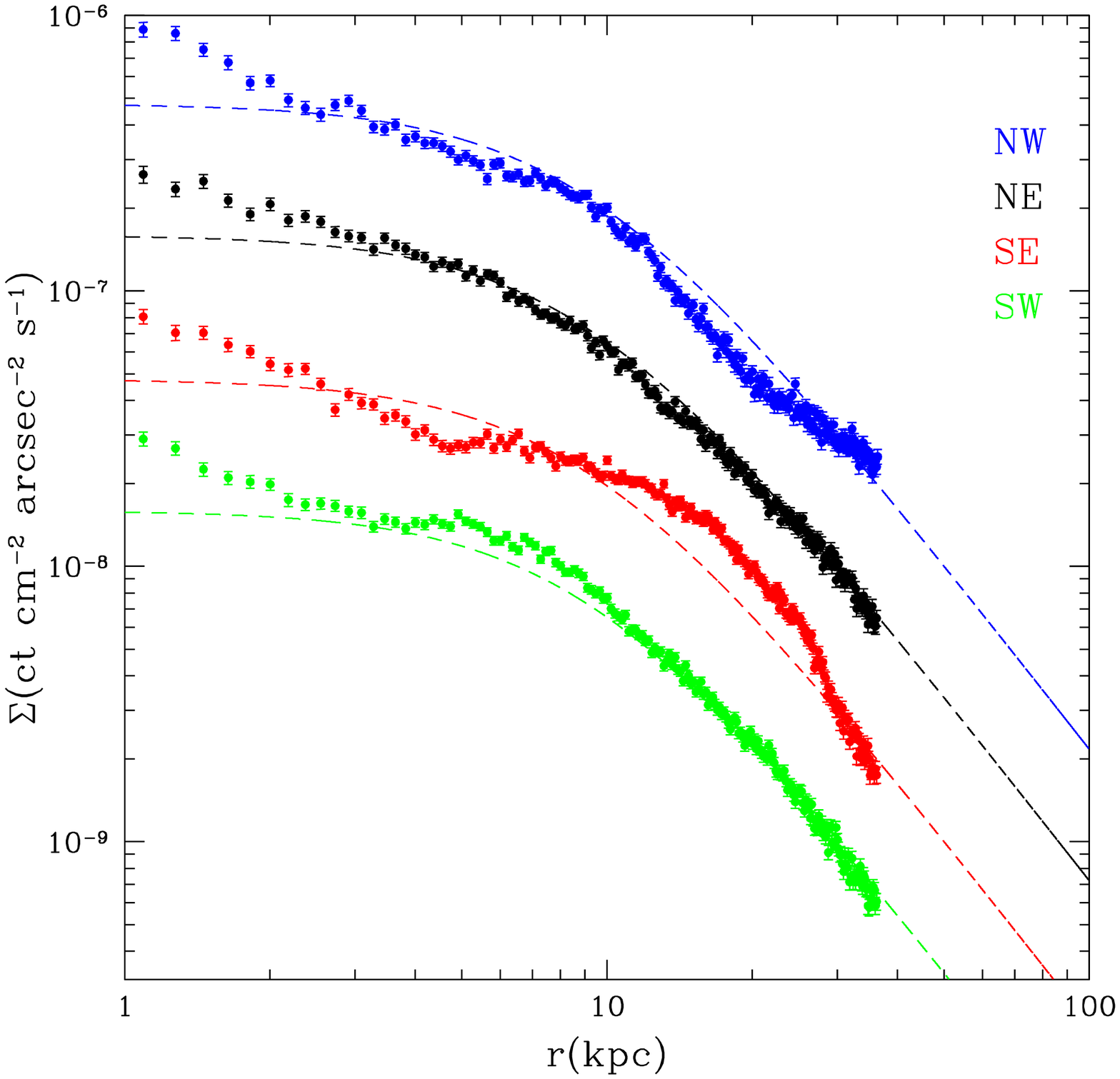}}
\caption{Background subtracted and exposure corrected 0.3-2.0~keV surface brightness profiles
in 4 quadrants. The profiles are off-set for clarity of presentation. For comparison, the
best-fit azimuthally average $\beta$ model (shown in Fig. 14) is plotted as a dashed line.}
\end{inlinefigure}

\noindent
Fig. 14 is in good agreement with 
the ROSAT PSPC analysis which included emission out to $40^{\prime}$ 
($\beta=0.53$; David et al. 2001) and the XMM-Newton PN surface brightness profile 
fitted by Buote et al. (2003), who obtained $\beta=0.52$.

\subsection{Azimuthal Variations in the Surface Brightness Profile}

We also generated background-subtracted and exposure corrected surface brightness 
profiles in the 0.3-2.0~keV energy 
band in 4 quadrants to a projected radius of 40~kpc, which is the 
full extent of the S3 chip (see Fig. 15).  This figure shows that 
there are significant azimuthal variations in the surface brightness profile within 
the central 30~kpc (the location of the SE cold front).  Beyond the SE cold front,
all 4 surface brightness profiles are similar.  The inner cavities show up 
as depressions 
in the surface brightness profiles within the central 10~kpc.  
Between 10 and 30~kpc, the surface brightness profiles toward the north-east and south-west
(perpendicular to the direction of the SE cold front) closely follow the 
azimuthally averaged surface brightness profile.  The south-east surface brightness 
profile is significantly higher than the azimuthally averaged value 
between 10-30~kpc, 
while the north-west surface brightness profile is significantly lower in this region.  
The SE cold front is clearly visible
as a break in south-east surface brightness profile at approximately 30~kpc.
There is also a break in the north-west surface brightness profile at 30~kpc in
the opposite sense (i.e., shallow inner surface brightness and steep
outer surface brightness).  The azimuthal variations in the surface brightness
profiles between 10 and 30~kpc are probably due to the sloshing motion of
NGC 5044 toward the south-east.

To further search for any features in the ACIS image of NGC 5044, we computed surface
brightness profiles within several concentric annuli as a function of position angle
(see Fig. 16).  
The SE spiral shows up as a spike in the surface brightness 
profiles at a position angle of approximately 140$^{\circ}$ and is clearly evident 
to the edge of the S3 chip. There is also a smaller amplitude spike in all 4 surface brightness 
profiles shown in 

\begin{inlinefigure}
\center{\includegraphics*[width=1.00\linewidth,bb=20 145 569 695,clip]{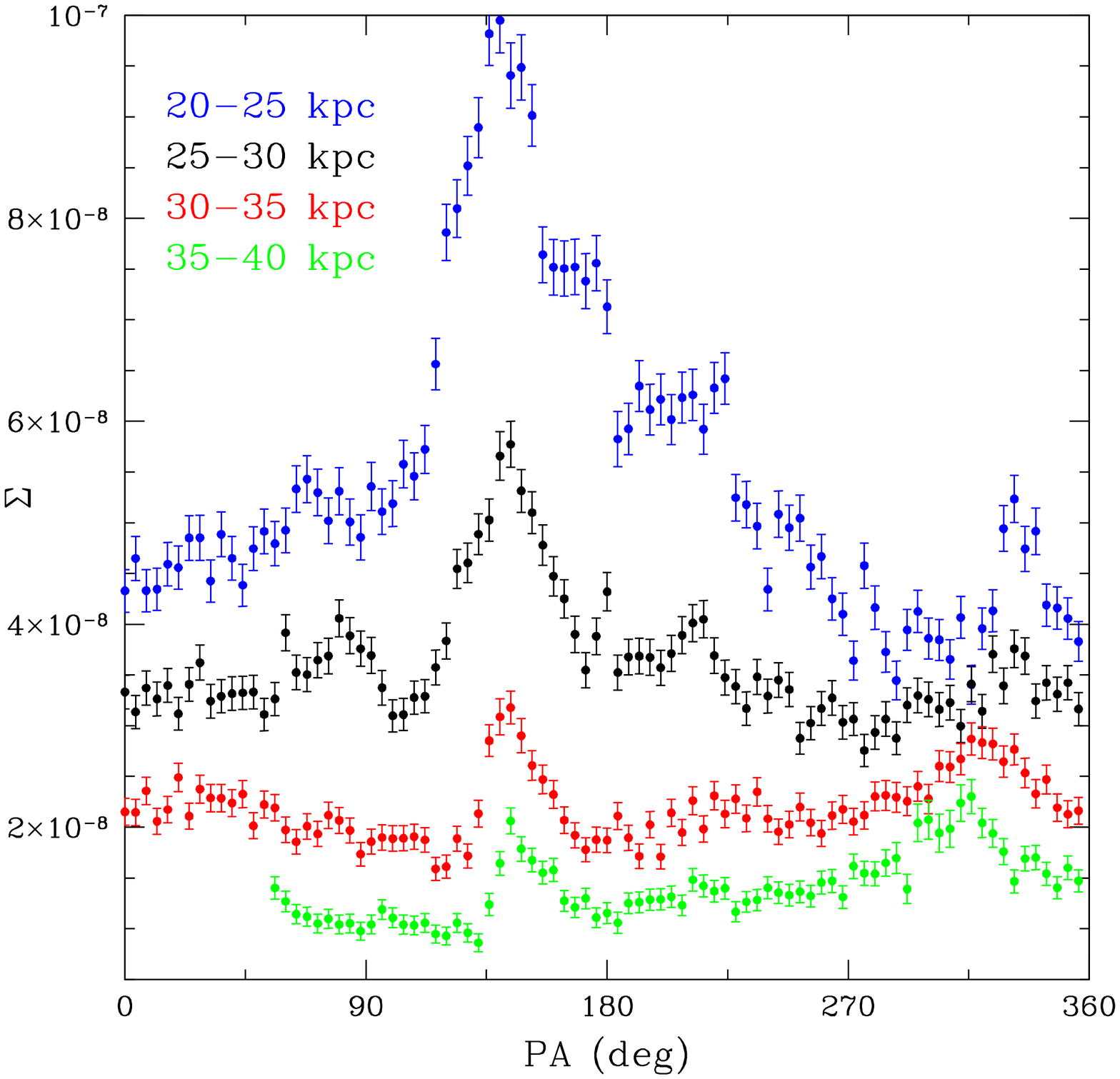}}
\caption{Background subtracted and exposure corrected 0.3-2.0~keV surface brightness profiles
within several annular regions vs. position angle (measured counter clockwise from due north).}
\end{inlinefigure}

\begin{inlinefigure}
\center{\includegraphics*[width=1.00\linewidth,bb=20 145 569 695,clip]{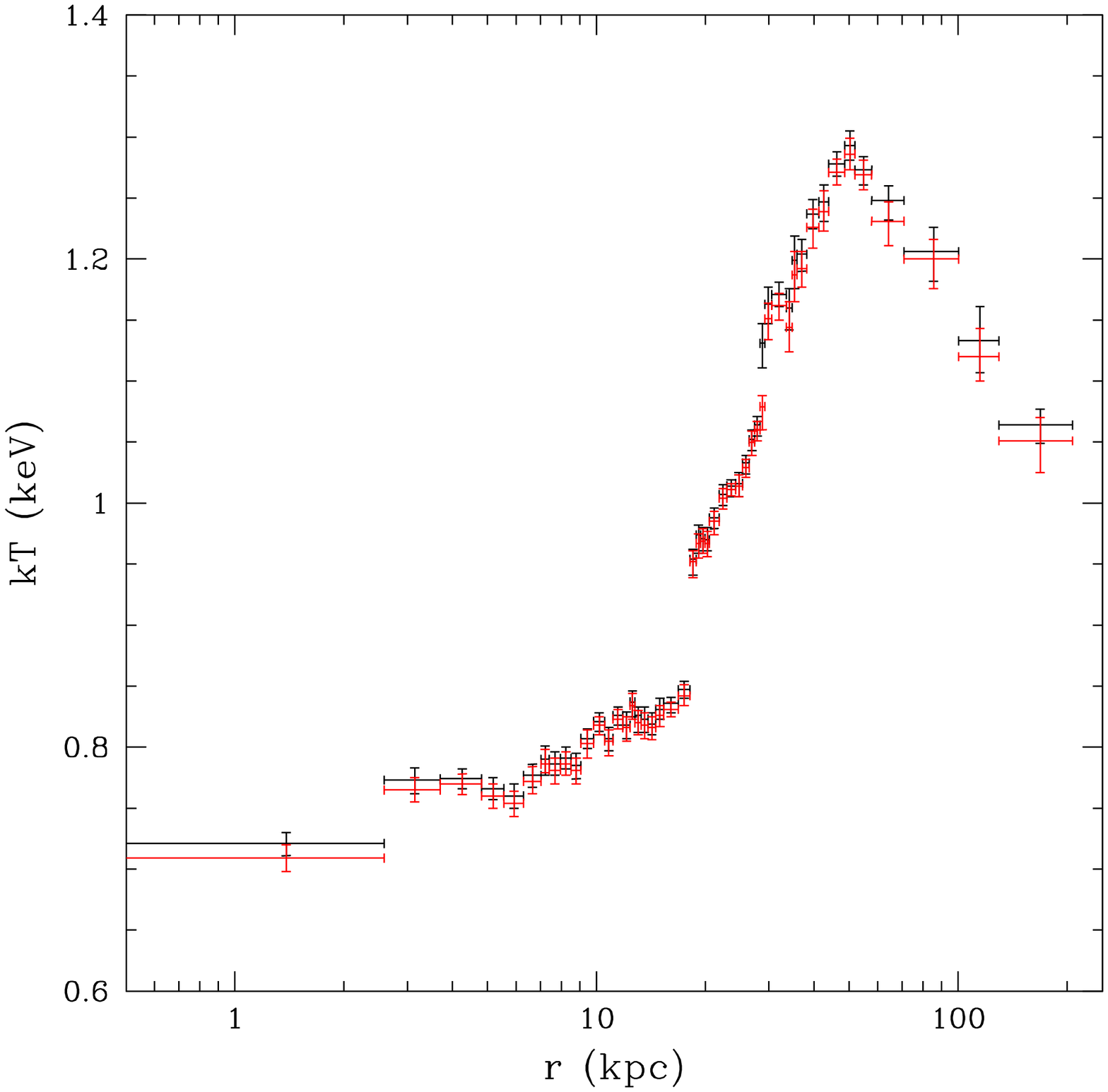}}
\caption{Azimuthally averaged projected temperature profile.  The temperatures are derived
by fitting the spectra in the 0.5-5.0~keV energy band to an absorbed single temperature
model (black data points) and an absorbed single temperature plus power-law model (red data points).}
\end{inlinefigure}

\noindent
Fig. 16 at a position angle of approximately 325$^{\circ}$.  
This NW spiral is only marginally visible in the raw image shown in Fig. 2, but is quite 
evident in Fig. 16.

\section{Projected Gas Temperature Profiles}

Based on the surface brightness profile, we extracted 
a set of spectra within concentric annuli with 10,000 net counts per
spectrum. The total net counts are based on the sum of the counts on all 5 chips. 
Source spectra, background spectra, photon weighted response files and photon
weighted effective area 

\begin{inlinefigure}
\center{\includegraphics*[width=1.00\linewidth,bb=20 145 569 695,clip]{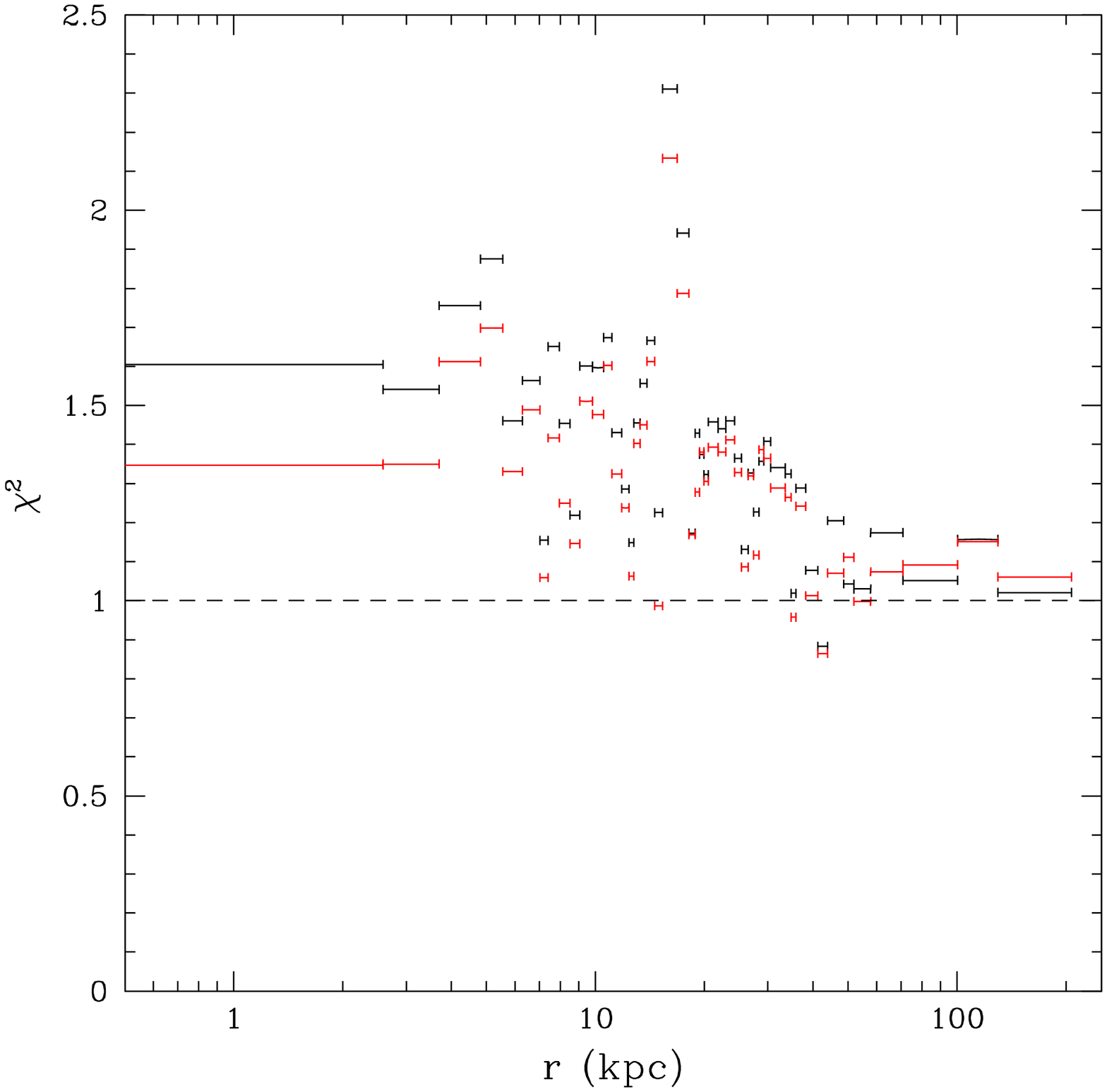}}
\caption{Reduced $\chi^2$ for the azimuthally averaged spectra fit
to an absorbed single temperature model (black data points) and an absorbed single temperature
plus power-law model (red data points).}
\end{inlinefigure}

\noindent
files were generated for each region after
excising the emission from point sources (i.e., point sources detected
by {\it wavdetect} in the 0.3-2.0 and 2.0-7.0~keV energy bands) using the CIAO 
task {\it specextract}.  The background spectra are extracted from the 
normalized blank sky data sets discussed in $\S 2$.  Beyond $200^{\prime\prime}$ 
(the region fully covered by the S3 chip), the spectra from
different chips are 
fit simultaneously.  Each spectrum was fitted in the 0.5-5.0~keV energy band to 
an absorbed vapec thermal plasma model and an absorbed vapec plus power-law model 
with the index frozen at $\Gamma=1.4$ to account for the emission from unresolved 
LMXBs.  The hydrogen column density was fixed to 
the galactic value, $N_H = 4.94 \times 10^{20}$~cm$^{-2}$.  We adopt the abundance table of 
Grevesse \& Sauval (1998) in our spectral analysis and treat O, Ne, Mg, Si,
S, Ar and Fe as free parameters with the Ni abundance linked to the 
Fe abundance and the Ca abundance linked to the Ar abundance.

\subsection{Azimuthally Averaged Projected Temperature profile}

From our moderately deep Chandra observation, we are able to derive a high spatial
resolution temperature profile of NGC 5044 (see Fig. 17).  The temperatures derived
from a single 
thermal plasma model and a thermal plus power-law model are
fully consistent (compare the black and red data points in Fig. 17).  
Within the 0.5-5.0~keV energy band, 85\% of the detected
photons have energies below 1.5~keV for 1~keV gas, so the spectral fits
are heavily weighted by the soft thermal emission and are not significantly 
affected by the harder emission from unresolved LMXBs.
The overall temperature profile shown in Fig. 17 is
similar to that observed in other groups with a positive temperature 
gradient at small radii, a peak temperature at approximately 50~kpc, and a 
declining temperature profile at larger radii (e.g., Finoguenov et al. 2007; 
O'Sullivan et al. 2009).  Within the central 10~kpc, the
azimuthally averaged temperature is nearly isothermal with temperatures
between 0.70-0.85~keV (consistent with the temperature map shown in Fig. 9).
There are two sharp temperature jumps at 20 and 30~kpc

\begin{inlinefigure}
\center{\includegraphics*[width=1.00\linewidth,bb=20 145 569 695,clip]{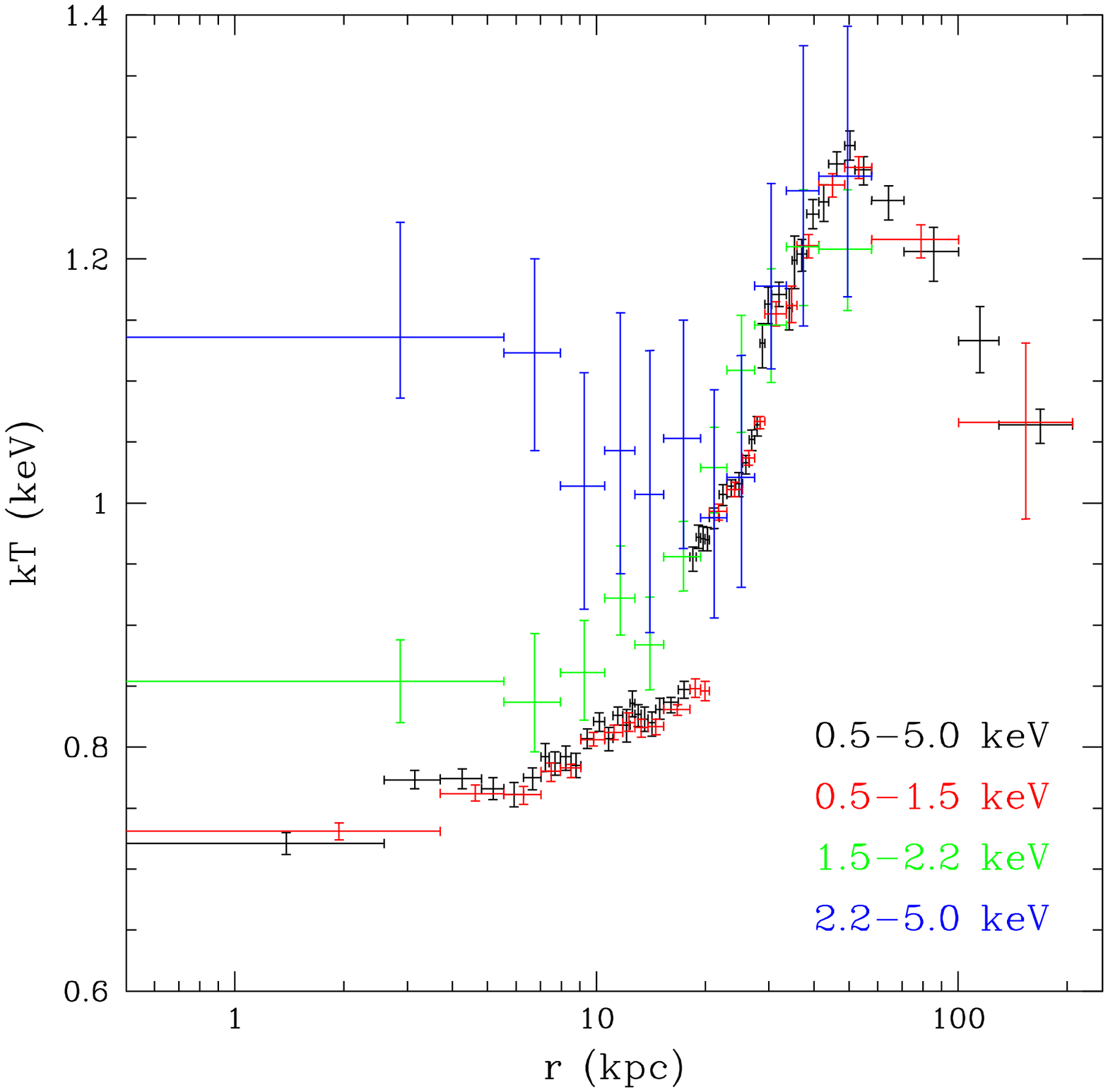}}
\caption{Azimuthally averaged projected temperature profile derived from fitting the spectra
in 4 different energy bands: 1) a broad energy
band from 0.5-5.0, 2) a soft energy band from 0.5-1.5~keV which contains mostly emission
from Fe-L lines, 3) an energy band from 1.5-2.2~keV which contains
line emission from H-like and He-line Si, and 4) a hard energy band from 2.2-5.0~keV
which contains some continuum plus line emission from S, Ar and Ca. }
\end{inlinefigure}

\noindent
seen in Fig. 17.  
The outer temperature jump
corresponds to the SE cold front.  The inner temperature
jump is primarily due to the non-azimuthal structure of the central
gas temperature (see the temperature map in Fig. 9 and $\S 6.2$).  Within 20~kpc, the 
spherical annuli only contain emission from the coolest gas.
Beyond this radius, the spherical annuli also contain some emission
from hotter gas.  The reduced $\chi^2$ values for the 
spectral fits are shown in Fig. 18.  
While the inclusion of a power-law component
in the spectral analysis does not affect the derived gas temperatures, it does
significantly reduce the $\chi^2$ value in most spectra.
For example, applying a F-test on the spectral fitting results 
of the innermost 5 annuli shows that the improvement in the fits
is significant at more than 98\% in each case.
Fig. 18 shows that single temperature fits are reasonably acceptable beyond 50~kpc, which 
corresponds to the peak of the temperature profile and the extent of the cool 
SE spiral.  The degradation in the spectral fits within this region is primarily 
due to projection effects, azimuthal variations in the gas temperature 
and the presence of multi-phase gas, as can be seen in the temperature map shown in Fig. 9.    

We also fitted the azimuthally averaged spectra to an absorbed thermal plasma 
model in three additional energy bands to determine the influence of the emission 
from unresolved LMXBs: 1) a soft energy band from 0.5-1.5~keV which contains mostly Fe-L
emission lines plus some weaker emission 
from O, Ne and Mg, 2) 
an energy band from 1.5-2.2~keV, which contains emission lines from He-like
and H-like Si, and 3) a harder energy band from 2.2-5.0~keV, which contains
mostly continuum emission plus emission lines from S, Ar and Ca (see Fig. 19).
For the soft energy band we used a set of spectra with 20,000 net counts
each and for the harder energy bands we used a set of spectra
with 40,000 net counts each.
Since most of the photons in the 0.5-5.0~keV energy band have energies below 1.5~keV,
the temperature profile derived in the broad and soft energy bands are in good
agreement.  The temperature profiles derived from the Si line ratio 
and the hard energy band also agree with the temperature profile derived from the 
broad and soft energy bands beyond 30~kpc.   The influence of the LMXBs on the 
derived gas temperatures is most notable in the hard energy band within the central 30~kpc.

To unambiguously determine if the harder spectral component in the center of 
NGC 5044 arises from an extra component of hotter gas or LMXBs, we fitted the 
1.5-7.0~keV emission 
within the central 8~kpc to a single temperature model, 
a two-temperature model (with the abundances linked) and a single temperature 
plus power-law model (see the results in Table 1).  
As can be seen in Table 1, the thermal plus power-law model provides the
best fit to the emission within the central 8~kpc.  In addition, the best-fit 
temperature is consistent with the central gas temperature derived above and 
the best-fit power-law index is consistent with the characteristic power-law 
emission from LMXBs.  The unabsorbed 2.0-7.0~keV flux of the power-law
component is F(2-7~keV)$=6.6 \times 10^{-14}$~erg~cm$^{-2}$~s$^{-1}$.  
Combining this with the 2.0-7.0~keV flux of the resolved point sources within the central
8~kpc gives a total 2.0-7.0~keV flux and luminosity for the LMXBs of
F(2-7~keV)$=7.5 \times 10^{-14}$~erg~cm$^{-2}$~s$^{-1}$  
and L(2-7~keV)$=4.5 \times 10^{39}$~erg~s$^{-1}$.
The 2.0-7.0~keV flux from the LMXBs within the central 8~kpc is
17\% of the total flux from the galaxy.
In David et al. (2006), we showed that there is a strong correlation between 
the total 0.5-2.0~keV luminosity of the LMXBs in 
early-type galaxies and the K-band 
luminosity of the galaxies.  An analysis of the 2MASS K-band image of NGC 5044 gives 
$\rm{m_K}=8.00$ and $\rm{L_{K}}=1.96 \times 10^{11} \rm{L_{K\odot}}$ within the 
central 8~kpc.  Using eq. (1) in David et al. (2006), we find a predicted luminosity of 
L(0.5-2.0~keV)=$4.1 \times 10^{39}$~erg~s$^{-1}$.  A power-law spectral 
model with $\Gamma=1.4$ gives L(2-7~keV)=$8.1 \times 10^{39}$, which is consistent 
with the observed 2-7~keV luminosity, given the observed scatter in the relationship.
This analysis shows that the harder X-ray component in the center of NGC 5044 is most 
likely due to unresolved LMXBs and not a second component of hotter gas.  
In general, since the X-ray emitting gas has a more extended distribution than the 
stars and LMXBs in early-type galaxies and the gas is the coolest in the center, the 
harder X-ray emission from LMXBs should be the most prevalent in the center of these galaxies.

\subsection{Azimuthal Variations in Projected Temperature Profiles}

The projected temperature profiles within 4 quadrants out to 40~kpc (the
field of view of the S3 chip) are shown in Fig. 20.  The SE cold front 
shows up as a sharp temperature jump in the south-east temperature profile at a radius
of 30~kpc.  Interior to the SE cold front, the gas is essentially isothermal.
The north-west temperature profile shows a temperature jump at 20~kpc which 
coincides with the break in the north-west surface brightness profile (see Fig. 15).  
There is a also a significant temperature jump in the south-west temperature profile 
at a radius of 20~kpc, but there is no obvious break in the surface brightness at this 
location (see Fig. 15).  This temperature jump is likely due to non-azimuthal
variations in the gas temperature within the south-west quadrant. 
A further analysis of the temperature map shows that the temperature varies by 15\% at
a radius of 20~kpc within the SW quadrant (i.e., between position angles of 
180 and 270$^{\circ}$).  There are also weaker temperature jumps in 
the north-east and north-west directions at approximately 14~kpc.

~~~~~~~~~~~~~~~~~

\begin{table*}[t]
\begin{center}
\caption{Spectral Analysis of the Inner 8~kpc}
\begin{tabular}{lcccc}
\hline
Model & $\rm{kT_c}$ & $\rm{kT_h}$ & $\Gamma$ & $\chi^2/DOF$ \\
& (keV) & (kev) & & \\
\hline\hline
WABS*VAPEC          & 0.90 (0.87-0.93) &   -               &   -       &  167/124 \\
WABS*(VAPEC+VAPEC)  & 0.55 (0.45-0.67) & 1.26 (1.14-1.46)  &   -       &  152/122 \\
WABS*(VAPEC+POW)    & 0.83 (0.79-0.89) &   -               & 1.27 (0.79-1.87)&  146/122 \\
\hline
\end{tabular}
\end{center}
\noindent
\noindent
Notes: Spectral analysis of the emission from the central 8~kpc.  The spectrum was
fitted to a single temperature model, a two-temperature model and a single temperature 
plus power-law model.  The best-fit temperature in the single temperature model and 
the best-fit lower temperature in the two-temperature model are given by 
$\rm{kT_c}$. The best-fit higher temperature in the two-temperature fit is given by 
$\rm{kT_h}$. The best-fit index in the single temperature plus power-law fit
is given by $\Gamma$.  The last column gives the $\chi^2$ per degree of freedom.
All error bars are given at the 90\% confidence limit.
\end{table*}

\noindent
These jumps are also probably 
due to non-azimuthal temperature variations within these quadrants.

The presence of cool gas in the SE and NW spirals can best be shown by 
plotting the azimuthal variations in gas temperature within concentric annuli 
between 20 and 40~kpc from the center of NGC 5044 (see Fig. 21).  
This figure shows two sharp temperature drops at position angles
of 135$^{\circ}$ and 315$^{\circ}$, which are coincident with the SE and NW
spirals seen in the surface brightness profiles shown in Fig 16.
Within the inner annulus, there is also a broader dip in gas temperature between 
PA=0 and PA=270$^{\circ}$, due to the greater extent of cool gas in the south-east 
direction (see the temperature map in Fig. 9).

\section{Spectral Analysis of the AGN, Cavities and Filaments}

\subsection{The AGN and Innermost Gas}

To deduce the spectral properties of the AGN in NGC 5044, we extracted 
a spectrum from within a $1.2^{\prime\prime}$ (220~pc) radius region
centered on the AGN. Table 2 shows that the spectrum is poorly fitted with a 
thermal or power-law model, but is well fitted with a thermal plus power-law
model.  The best-fit temperature in the thermal plus power-law model is consistent
with the central gas temperature shown in the temperature map (see Fig. 9) and 
temperature profile (see Fig. 17).  We also allowed the hydrogen column density to vary
in the fitting process, but there is no evidence for excess absorption above 
the galactic value.  The best-fit power-law index in the two component model
is also quite typical of AGN. Based on the best-fit thermal 
plus power-law model, the unabsorbed 0.5-5.0~keV flux and luminosity of the AGN are
$1.98 \times 10^{-14}$~erg~s$^{-1}$~cm$^{-2}$ and $3.57 \times 10^{39}$~erg~s$^{-1}$
and the bolometric X-ray luminosity is 
$6.64 \times 10^{39}$~erg~s$^{-1}$.
The expected bolometric X-ray luminosity of the LMXBs based on the K-band 
luminosity of the galaxy 
within the central 220~pc is $3.3 \times 10^{37}$~erg~s$^{-1}$, 
which is less than 1\% of the total luminosity of the power-law component 
within this region.  The central radio point source was detected in our GMRT
observations at frequencies of 235~MHz, 327~MHz and 610~MHz, with flux densities 
of $35 \pm 2$~mJy, $32 \pm 2$~mJy and $30 \pm 2$~mJy, 
respectively (Giacintucci et al. 2009).  Integrating the radio flux density
between 5~MHz and 10~GHz, assuming $S_{\nu}\propto \nu^{-\alpha}$ with $\alpha=0.16$
gives a radio luminosity of $L_R=4.1 \times 10^{38}$~erg~s$^{-1}$.  This shows
that the radiative efficiency of the AGN in the radio band is only 6\% of that in the 
X-ray band.  Using a stellar velocity dispersion of $\sigma_* = 237$~km~s$^{-1}$ for NGC 5044 (which is
the average of 4 measurements listed in HyperLeda) and the relation
between stellar velocity dispersion and central black hole mass 
in Gebhardt et al. (2000), gives a

\begin{inlinefigure}
\center{\includegraphics*[width=1.00\linewidth,bb=20 145 569 695,clip]{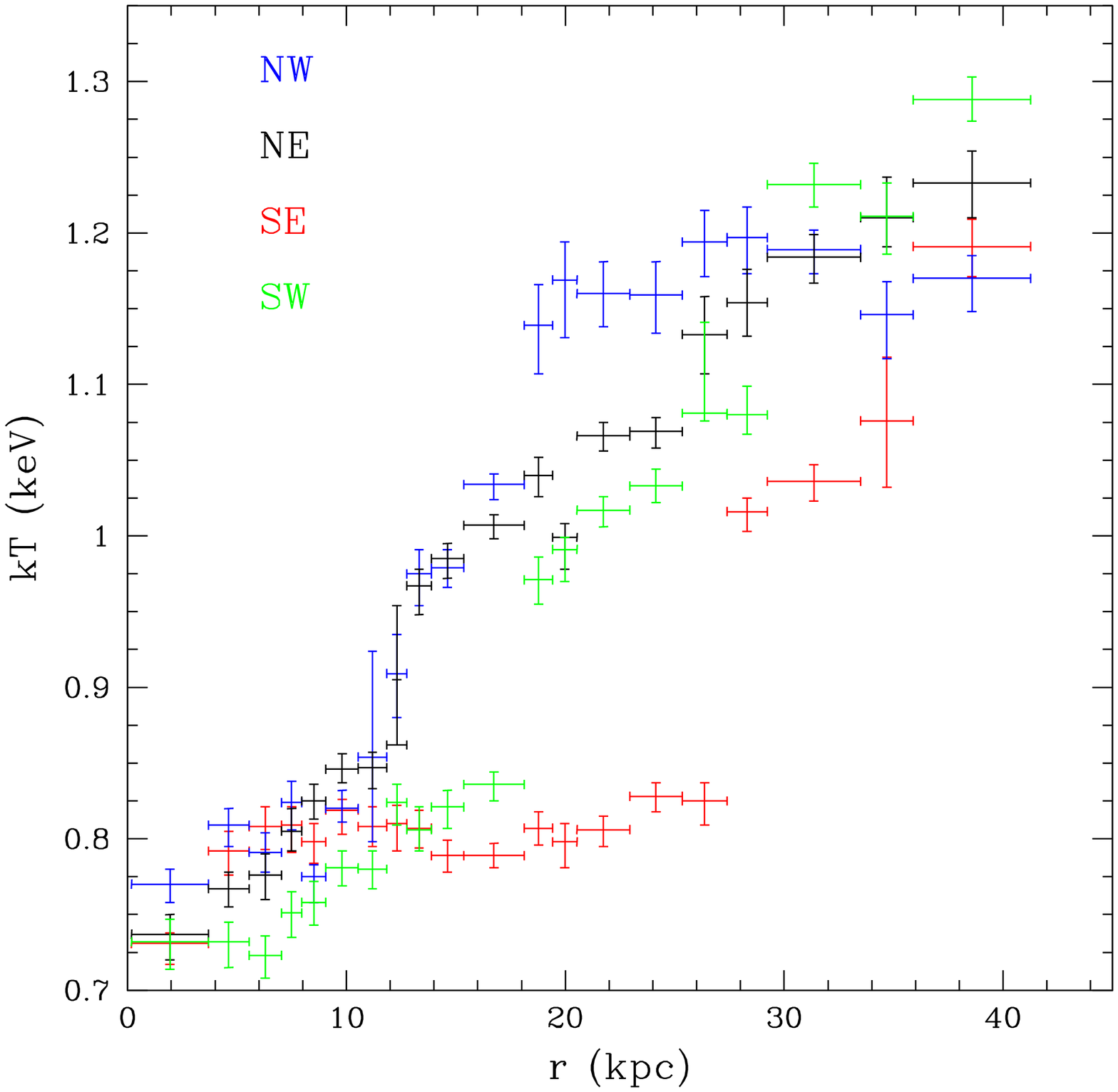}}
\caption{Projected temperature profile within 4 quadrants.  The temperatures are derived
by fitting the spectra in the 0.5-5.0~keV energy band to an absorbed single temperature
plus power-law model.}
\end{inlinefigure}

\noindent
central black hole mass of 
$\rm{M_{bh}}=2.27 \times 10^{8}~\Mo$.  This gives a ratio between the bolometric and 
Eddington luminosity of $\rm{L_{bol}/L_{Edd}} = 2.3 \times 10^{-7}$. 

The bolometric X-ray luminosity of the gas within the central 220~pc is
$1.56 \times 10^{39}$~erg~s$^{-1}$ based on the best-fit thermal plus 
power-law spectral model.  Using the best-fit 
emission measure of the thermal component,
assuming a uniform gas density within the central 220~pc, gives a density of $\rm{n_e=0.21}$~cm$^{-3}$ 
and a total gas mass of $\rm{M_{gas}}= 2.64 \times 10^{5}~\Mo$.
The isobaric cooling time of the gas is 
$\rm{t_c = 5kTM_{gas}/(2 \mu m_p L_{bol}})= 3.5 \times 10^7$~yr.
Alternatively, if we assume the gas density follows a power-law profile given 
by $\rm{n_e \propto r^{-0.8}}$, consistent with the density 
distribution determined
from a deprojected spectral analysis (David et al. 2009), 
we obtain $\rm{n_e = 0.14~(r/r_0)^{-0.8}}$~cm$^{-3}$, with $r_0=220$~kpc.

The Bondi accretion radius of the central black hole is
$\rm{R_a = G M_{bh}/c_a^2}$, where $\rm{c_a}$ is the adiabatic sound speed at the
accretion radius.  Using a gas temperature of 0.75~keV gives $\rm{R_a=4.9}$~pc.  
The Bondi accretion rate is given by $\rm{\dot M_B = 4 \pi R_a^2 \rho(R_a) c_a}$.
The greatest uncertainty in this

~~~~~~~~~~~~~~~~~~~~~~~~~~~~~~~~~

\begin{table*}[t]
\begin{center}
\caption{Spectral Analysis of the AGN}
\begin{tabular}{llccc}
\hline
Region & Model & kT & $\Gamma$ & $\chi^2/DOF$ \\
& & (keV) & \\
\hline\hline
AGN         & WABS*APEC         & 2.40 (1.39-3.34) & -                & 53.1/17 \\
AGN         & WABS*POW          & -                & 1.95 (1.79-2.10) & 53.8/18 \\
AGN         & WABS*(APEC+POW)   & 0.83 (0.75-1.02) & 1.44 (1.04-1.74) & 16.1/15 \\
\hline
\end{tabular}

\end{center}
\noindent
Notes: Spectral analysis of the emission from the central 1.2$^{\prime\prime}$ (220~pc) region.
The spectrum was fitted to a thermal plasma model, a power-law model and a thermal plus
power-law model.  The last column gives the $\chi^2$ per degree of freedom.
All error bars are given at the 90\% confidence limit.
\end{table*}

\begin{table*}[t]
\begin{center}
\caption{Spectral Analysis of the Filaments and Cavities}
\begin{tabular}{llcc}
\hline
Region & Model & kT & $\chi^2/DOF$ \\
& & (keV) & \\
\hline\hline
SE filament &  WABS*(APEC+POW)  & 0.83 (0.82-0.84) & 142/90  \\
SE filament &  WABS*(VAPEC+POW) & 0.79 (0.78-0.80) & 108/85  \\
SW filament &  WABS*(APEC+POW)  & 0.70 (0.68-0.71) & 89.0/75  \\
SW filament &  WABS*(VAPEC+POW) & 0.65 (0.64-0.66) & 69.3/70 \\
            &             &                  &          \\
NE cavity   &  WABS*(APEC+POW)  & 0.81 (0.80-0.82) & 72.3/65 \\
NE cavity   &  WABS*(VAPEC+POW) & 0.81 (0.79-0.83) & 59.1/60 \\
S  cavity   &  WABS*(APEC+POW)  & 0.84 (0.83-0.85) & 139/88  \\
S  cavity   &  WABS*(VAPEC+POW) & 0.82 (0.80-0.84) & 114/83  \\
SW cavity   &  WABS*(APEC+POW)  & 0.78 (0.77-0.79) & 68.6/70  \\
SW cavity   &  WABS*(VAPEC+POW) & 0.73 (0.71-0.75) & 42.1/65  \\
NW cavity   &  WABS*(APEC+POW)  & 0.82 (0.81-0.84) & 142/74  \\
NW cavity   &  WABS*(VAPEC+POW) & 0.78 (0.76-0.80) & 87.1/69 \\
\hline
\end{tabular}
\end{center}
\noindent
Notes: Spectral analysis of the filaments and cavities shown in Fig. 22.
The last column gives the $\chi^2$ per degree of freedom.
All error bars are given at the 90\% confidence limit.
\end{table*}

\begin{inlinefigure}
\center{\includegraphics*[width=1.00\linewidth,bb=20 145 569 695,clip]{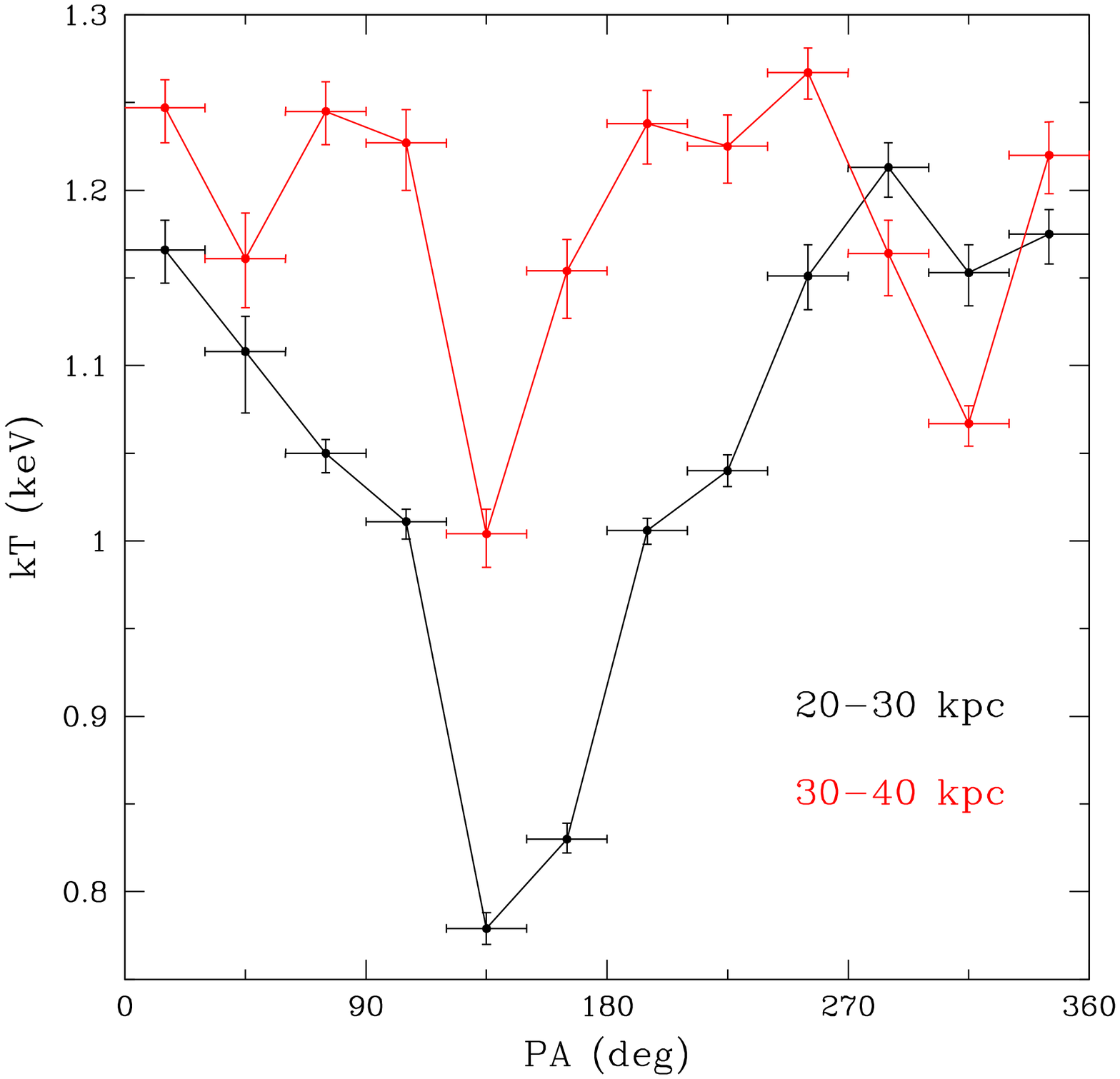}}
\caption{Projected temperature profile within two annular regions vs. position angle
(measured counter clockwise from due north).}
\end{inlinefigure}

\noindent
expression is the gas density at the Bondi radius.
Even in the nearby NGC 5044 group, an extrapolation over a factor of 40 in radius
is required
to estimate the density at $\rm{R_a}$. If the gas is isothermal and the
density follows the same power-law expression given above, then
$\rm{\dot M_B = 0.011~\Mo}$~yr$^{-1}$.  
If the X-ray luminosity of the AGN is powered by Bondi accretion, the resulting efficiency 
($\rm{\epsilon = L_{bol}/(\dot M_B c^2}$) is $1.1 \times 10^{-5}$, which is typical of 
the radiatively inefficient AGN in early-type galaxies (Loewenstein et al. 2001; 
Di Matteo et al. 2003; David et al. 2005).

\vskip 0.5in

\subsection{Cavities and Filaments}

Spectra were extracted from the cavities and filaments using the regions
shown in Fig. 22 and fitted to absorbed apec and vapec models in the 0.5-5.0~keV
energy band.  A power-law component was added in both cases to account for the 
emission from unresolved X-ray binaries with the power-law index
frozen at $\Gamma=1.4$.  Table 3 shows that 
the apec model provides a poor fit to the spectra extracted from both 
the cavities and filaments. A significant improvement is obtained with the vapec 
model indicating that either the abundance ratios differ from those
in the abundance table of Grevesse \& Sauval (1998), or the gas
is multi-phase (especially as seen in projection), 

\begin{inlinefigure}
\includegraphics*[width=1.00\linewidth,bb=102 248 436 544,clip]{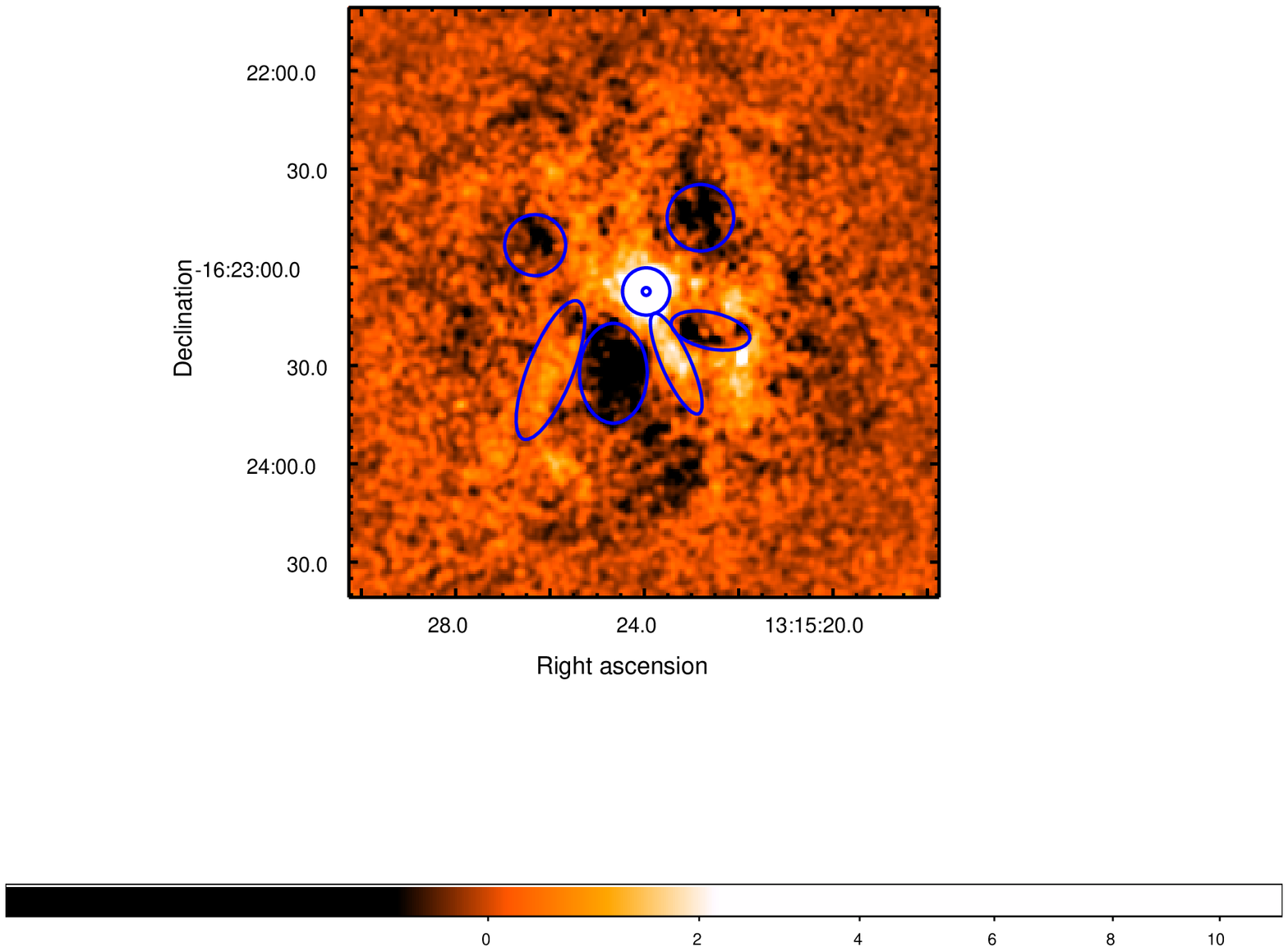}
\caption{Regions used to extract spectra for the cavities, AGN and filaments (see Table 3). }
\end{inlinefigure}

\noindent
and that different lines are 
being excited by different temperature gas.  However, as is also evident from the 
temperature map (in which the gas temperature is derived solely from the 
centroid of the Fe-L lines), the cavities have higher temperatures than the 
filaments, but only by 0.05-0.1~keV.  Both the temperature map and temperature 
profile show that the gas is nearly isothermal within the central 10~kpc.  
Most of the cavities are only a few kpc in extent, so the gas that was 
displaced likely had a temperature similar to the surrounding gas seen in projection.
The coolest gas within the central region of NGC 5044 is located within the 
SW filament with $\rm{kT} = 0.65$~keV (see Table 3).   This is also evident
in the temperature map derived from the earlier
20~ksec ACIS observation presented by Gastaldello et al. (2008).

We also searched for evidence of hot thermal gas entrained within the 
cavities by extracting a spectrum of the combined emission from all
4 cavities shown in Fig. 22.  The extracted spectrum shows no evidence 
for any Fe-K$\alpha$ emission, which 
would be an unambiguous sign of shocked hot gas in a group of galaxies.  
The net count rate between 5 and 7~keV is actually consistent with zero 
with a $3 \sigma$ upper limit of $1.1 \times 10^{-4}$~ct~s$^{-1}$.  
If we scale the expected emission from X-ray binaries according to the
K-band luminosity of the galaxy within the 4 cavities we find
a predicted 5-7~keV count rate from X-ray binaries of $1.0 \times 10^{-4}$~ct~s$^{-1}$,
which is just below our $3 \sigma$ upper limit. 
Since the projected cavity spectrum contains thermal
emission from gas with temperatures between 0.8 and 1.4~keV, and
since we are only interested in detecting gas hotter than that
encountered along the line-of-sight, we fitted the spectrum in the 2.5-5.0~keV 
energy band to an absorbed vapec plus power-law model (to account
for the emission from unresolved LMXBs) and obtained a best-fit temperature of 
kT=$0.90 \pm 0.30$~keV (90\% errors), which is consistent with emission 
from ambient gas along the line-of-sight.  

To place constraints on the amount of hot gas entrained within the
cavities, we computed the predicted count rate in the 5-7~keV energy band 
for gas in pressure equilibrium with the surrounding medium assuming a solar 
abundance of Fe (see Table 4).  This table shows that the predicted count rates for
gas hotter than 2~keV are all well below the $3 \sigma$ upper limit and the count 
rate expected from unresolved LMXBs.  Thus, we cannot exclude the presence of gas with 
temperatures as low as 2~keV in pressure equilibrium within these cavities.

\section{Discussion}

\subsection{Interaction of the Radio Plasma with the Hot Gas}

NGC 5044 hosts many small X-ray cavities with a nearly isotropic distribution
and one larger cavity toward the south. The smaller cavities are
all radio quiet, even at low frequencies (sensitivity level
of 0.25~mJy/b at 235~MHz; Giacintucci et al. 2009b).
This suggests that these cavities are no longer powered by the central
AGN and are primarily driven by buoyancy.  The isotropic distribution of the
small, buoyantly driven cavities may be due to group "weather" as in 
the simulations of 
Bruggen et al. (2005) and Heinz et al. (2006).  The origin 
of the group weather could be due to previous AGN outbursts or the 
sloshing of NGC 5044 with respect to the center of the group potential.

The enthalpy in a cavity is given by $H = \gamma~PV/(\gamma-1)$ and varies between
2.5~PV and 4~PV for $\gamma=5/3$ and $\gamma=4/3$. There are many uncertainties
in estimating the ages of a cavities and McNamara \& Nulsen (2007)
list three possible methods for computing their ages, including: the
sound crossing time, the buoyancy time and the refill time.  We computed 
the three different age estimates for each cavity and found that the age
estimates agree to within 30\%.  As a reasonable approximation, we compute the cavity 
power as $P_c=4PV/t_s$, where $t_s$ is the sound crossing time between the AGN and the 
center of the cavity.  This gives a total cavity power for the three small 
radio quiet cavities (i.e., the SW, NW and NE cavities) 
of $9.2 \times 10^{41}$~erg~s$^{-1}$.  The cavity power of the larger southern
cavity is $5.6 \times 10^{42}$~erg~s$^{-1}$.  
For comparison, the bolometric luminosity of the hot gas within the central 
10, 20 and 30~kpc is $2.1 \times 10^{42}$~erg~s$^{-1}$, 
$4.7 \times 10^{42}$~erg~s$^{-1}$ and 
$7.1 \times 10^{42}$~erg~s$^{-1}$, respectively.
This shows that the cavity power of the three smaller cavities,
currently located between 4 and 7~kpc from the center of NGC 5044,
can balance about one-half of the radiative losses within 
the central 10~kpc.  However, the combined cavity power of the three small
cavities plus the larger southern cavity is sufficient to balance all radiative
losses within the central 25~kpc. 
As discussed below, the presence of $H\alpha$ emitting gas within the 
central 5~kpc shows that as least some of the hot gas is able to 
cool to low temperatures.

Studies have shown that the mechanical cavity power far exceeds the 
radio luminosity of the X-ray cavities in groups and clusters of galaxies 
and that the ratio of cavity power to radio luminosity increases with
decreasing radio luminosity (Birzan et al. 2004; 2008).  In this paper we
concentrate on the properties of the radio lobe toward the south-east
observed at 610~MHz.  The analysis of the extended radio emission at 235~MHz 
within the southern cavity is more complicated due to the larger beam
size at 235~MHz (22$^{\prime\prime}$ by 16$^{\prime\prime}$) and 
the 20$^{\prime\prime}$ separation between the southern X-ray cavity
and the central AGN.  The GMRT data at 235~MHz will be discussed in more
detail in  Giacintucci et al. 2009b).

The monochromatic 610~MHz luminosity of the south-east radio lobe in NGC 5044 is 
$L_{610}=9.3 \times 10^{36}$~erg~s$^{-1}$ (Giacintucci et al. 2009a).
Assuming a spectral index of $\alpha=1$ and integrating between 
10~MHz and 5~GHz (as in Birzan et al. 2004)
gives a radio luminosity of $L_R=5.7 \times 10^{37}$~erg~s$^{-1}$,
which is a factor of approximately 100 less than the least luminous radio filled cavity
studied in the Birzan et al. (2004) sample.  The ratio of cavity power to radio 
luminosity, assuming the cavity is the same size as the radio lobe, 
is $P_c/L_R = 1.3 \times 10^3$ for the south-east radio lobe in NGC 5044.  
Using the relation between 
cavity power and radio luminosity (eq. 12 in Birzan et al. 2008) along
with the observed radio luminosity of the south-east radio lobe, gives
a predicted cavity power of $P_c = 4.0 \times 10^{41}$, which is only 
40\% less than the observed value. While this is only one extra data point, 
it shows that there is some evidence that the relation between $P_c$ and $L_R$ 
derived in Birzan et al. (2008) extends over 8 orders of magnitude in $L_R$.

We showed above that we cannot place very tight constraints on the 
amount of hot thermal gas within the cavities in NGC 

\begin{table*}[t]
\begin{center}
\caption{Predicted Count Rates for Entrained Thermal Gas in the Cavities}
\begin{tabular}{lcc}
\hline
kT & $\rm{n_e}$ & R(5.0-7.0~keV) \\
(keV) & (cm$^{-3}$) & ct~s$^{-1}$ \\
\hline\hline
2.0  & $8.0 \times 10^{-3}$ & $1.61 \times 10^{-5}$ \\
3.0  & $5.3 \times 10^{-3}$ & $1.31 \times 10^{-5}$ \\
4.0  & $4.0 \times 10^{-3}$ & $9.97 \times 10^{-6}$ \\
5.0  & $3.2 \times 10^{-3}$ & $7.68 \times 10^{-6}$ \\
10.0 & $1.6 \times 10^{-3}$ & $2.45 \times 10^{-6}$ \\
\hline
\end{tabular}
\end{center}
\noindent
Notes: This table gives the predicted 5-7~keV count rate for thermal 
gas in pressure equilibrium within the 4 cavities assuming gas temperatures 
between 2 and 10~keV. The electron number density, $n_e$, is derived by
assuming pressure equilibrium.  The $3 \sigma$ upper limit on the observed 5-7~keV
count rate within these cavities is $1.1 \times 10^{-4}$~ct~s$^{-1}$.
\end{table*}

\noindent
5044 due to the 
hard X-ray emission from unresolved LMXBs.  Constraints on the particle
content within the radio lobes observed in groups and clusters can 
be obtained by assuming the lobes are in pressure equilibrium
with the ambient gas (Dunn \& Fabian 2004; Birzan et al. 2008; Croston et al. 2008).
The total particle energy within a radio lobe can be written as
$E_p = k E_e$, where $E_e$ is the energy in relativistic
electrons emitting synchrotron radiation at frequencies greater than 10~MHz.
Using the observed 610~MHz radio luminosity from the south-east radio lobe along
with the expression for $E_e$ in Pacholczyz (1970), assuming $\alpha=1$, gives:

$$E_p = 6.0 \times 10^{54} \left( { {B} \over {1~\mu G}} \right) k~\rm{erg}$$

\noindent
The energy in the magnetic field is $E_B = B^2\phi V / (8 \pi)$, where
V is the volume of the radio lobe and $\phi$ is the volume filling factor. 
For the south-east radio lobe, this gives:

$$E_B = 1.4 \times 10^{54} \left( { {B} \over {1~\mu G}} \right)^2 \phi ~\rm{erg}$$

\noindent
Assuming energy equipartition then gives:

$$B_{eq} = 1.5 \left( { {k} \over {\phi}} \right) \mu G$$ 

\noindent
Combining the assumptions of energy equipartition and pressure equilibrium 
between the ambient medium and the radio lobe
($P_T = k E_e/(3 V \phi) + E_B/ (V \phi)$)
gives $k/\phi=1.2 \times 10^4$ and $B_{eq}=23 \mu G$.


Performing a similar analysis, Dunn \& Fabian (2004) and Birzan et al. (2008)
obtained values of $k/\phi$ up to approximately 4000 based on samples
of radio lobes in mostly rich clusters of galaxies.  This comparison shows that 
the non-radiating energy content of the south-east radio lobe in NGC 5044 is 
significantly greater than that in the larger radio lobes analyzed in 
these two samples. Birzan et al. (2008) found a correlation between
$k/\phi$ and the synchrotron age of the radio lobe. Based on an analysis of 
radio lobes in a sample
of groups of galaxies, Croston et al. (2008) noted that "plumed" radio
lobes were more under-pressurized than radio lobes connected with a jet.  Both of 
these results suggests that $k/\phi$ increases with time after 
the radio lobes are no longer powered by the central AGN.  The most likely 
origins for this increase in $k/\phi$ are the entrainment of thermal gas 
and aging of the relativistic electron population.
While we cannot estimate the synchrotron age for the south-east radio lobe without
knowing the break frequency in the radio spectrum, the sound crossing time between the AGN 
and the south-east lobe is only $2 \times 10^7$~yr, so it must be fairly young.  However,
the torus-like structure of the south-east radio lobe and the fact that a cool filament of
gas is currently threading the torus suggest that it may have already entrained
a significant amount of ambient gas.

\subsection{Comparison with Optical and IR observations}

Most of the H${\alpha}$ emission from NGC 5044 is contained in a broad
filament aligned in a north to south direction (Macchetto et al. 1996).  
The northern portion of the H${\alpha}$ filament has an extent of 
approximately 7~kpc and lies along the eastern rim of the NW cavity. The southern 
portion of the filament trails the southern cavity, but unfortunately,
a "ghost image mutilated the H${\alpha}$+[NII] emission map" beyond approximately
5~kpc, which is only half of the distance to the center of the south-east radio torus.
Temi et al. (2007) showed that there is some evidence for distributed PAH emission 
with the same morphology as the southern extension of the H${\alpha}$ filament 
from the {\it Spitzer} observation of NGC 5044.  Spatial correlations between 
the H${\alpha}$ emission, PAH emission and X-ray morphology were also noted by 
Gastaldello et al. (2008) based on the earlier 20~ksec Chandra observation.

Caon et al. (2000) obtained long-slit spectroscopy at three position angles across
the H${\alpha}$ filament. The gas in the northern extension of the 
filament is mostly red-shifted with respect to NGC 5044 with a peak velocity of 
approximately 100~km~s$^{-1}$.  The kinematics of the gas in the
southern extension of the H${\alpha}$ filament are more chaotic with red
and blue shifted velocities between $\pm$~100~km~s$^{-1}$.  Without knowing the
inclination of the H${\alpha}$ filament relative to the plane of the sky, we 
cannot say which gas is in-falling and which gas is outflowing.  However, the 
kinematics of the gas in the H${\alpha}$ filament are very similar to 
the kinematics of the H${\alpha}$ filaments in NGC 1275 observed by Hatch et al.
(2006) and are probably due to dredge up and subsequent infall behind the 
buoyantly rising cavities.  Temi et al. (2007) also argued
that the extended dust distribution, as highlighted by the PAH emission,
could be due to the break up of a central dusty disk followed by buoyant uplifting
behind the southern X-ray cavity.

The kinematics of the warm $10^4$~K gas are inconsistent with an external 
origin for the gas (Caon et al. 2000), so the gas could arise from either stellar 
mass loss or radiative cooling of the hot ambient gas. 
The total mass of H${\alpha}$ emitting gas is $8.5 \times 10^5 \Mo$ (Macchetto et al. 1996).  
From the H${\alpha}$ image, it appears that most of the warm gas originates from 
within the central 2~kpc, and is then dredged outward.  Fitting a de Vaucouleurs
profile to the 2MASS K-band image of NGC 5044 gives a deprojected K-band luminosity within
the central 2~kpc of $L_K = 6.5 \times 10^9 \Lo$ (David et al. 2009).  Using the stellar 
mass loss rate in Athey et al. (2002) gives $\dot M_* = 0.012 \Mo$~yr$^{-1}$
within this region.
The cooling time of the hot gas within the central 2~kpc is $4.1 \times 10^7$~yr 
and the mass cooling rate is approximately $\dot M_h = 0.9 \Mo$~yr$^{-1}$. 
Thus, even if only 2\% of the hot gas is able to cool to low temperatures, 
radiative cooling of the hot gas would still be the dominant mass supply mechanism for 
the warm gas in NGC 5044.


\subsection{Residual Motion of NGC 5044}

NGC 5044 has two semi-circular cold fronts at projected distances of 30 
and 50~kpc from the center of the group. Cold fronts are commonly found near 
the central dominant galaxy in clusters and may be due to merger induced sloshing 
(Ascasibar \& Markevitch 2006; Markevitch \& Vikhlinin 2007).
Based on an optical study of the galaxies in the NGC 5044 group, Cellone \& Buzzoni 
(2005) found that NGC 5044 has a peculiar velocity of approximately 140~km~s$^{-1}$ 
with respect to the velocity centroid of the other galaxies in the group. 
One possible candidate for perturbing NGC 5044 is a sub-group of galaxies toward the NE
identified by Mendel et al. (2008), but this group lies at a projected distance
of 1.5~Mpc. Another possible candidate is the optically perturbed spiral galaxy
NGC 5054 which is located at a projected distance of 300~kpc
toward the south-east, has a blue-shifted velocity of 960~km~s$^{-1}$ and is
only two magnitudes fainter than NGC 5044.

Single spiral features have been observed in the X-ray images of the Ophiuchus cluster 
(Sanders and Fabian 2002), the Perseus cluster (Churazov et al. 2000; Fabian et al. 2006) 
and A2029 (Clarke et al. 2004).  Ascasibar \& Markevitch (2006) have shown through numerical 
simulations that single spiral arms of cool gas are commonly produced by sloshing 
of the central galaxy, however, none of their simulations produce two spiral arms
like those observed in NGC 5044.  The spiral arms in NGC 5044 are more similar to 
the spiral arms (or "hour glass" shaped feature) in NGC 4636
(Jones et al. 2002; Baldi et al. 2009).  X-ray spectral
analysis of the gas in the spiral arms in NGC 4636 shows that this gas 
was recently shocked, probably during the inflation of the radio emitting 
plasma currently observed between the spiral arms (Jones et al. 2002; Baldi et al. 2009).
The spiral arms in NGC 5044 are cooler than the surrounding gas.  
Also, no radio emission is detected at either 235~MHz or 610~MHz
interior to the spiral arms like that observed in NGC 4636.
Thus, if the spiral arms in NGC 5044 were created by the inflation of
radio bubbles, this must have occurred during an earlier radio outburst.  This
would also give the gas in the spiral arms time to cool by adiabatic expansion
if the were was initially shocked.

\section{Summary}

The Chandra observation of NGC 5044 provides us with some insights
on the AGN-cooling flow feedback mechanism that cannot be attained from 
observing more distant rich clusters. Unlike the large, energetic and nearly 
bi-polar distribution of the X-ray cavities observed in many rich clusters, 
the inner 5-10~kpc of NGC 5044 is filled with many small radio quiet cavities.  
Since these small cavities are probably no longer momentum driven by the central 
AGN, their motion will be driven by the group weather as they buoyantly rise outward.  
Hence, most of their energy will be deposited within the group center and isotropized 
by the group weather.  Such a situation 
could be a common feature in the center of groups and clusters, in which most of 
the central cooling and star formation is suppressed by many small, weather driven 
bubbles, while the less frequent larger outbursts predominately heat the gas at larger 
radii and supply some of the "pre-heating" required to break the 
self-similarity of groups and clusters.

We would like to thank T. Venturi and R. Athreya for help with the GMRT observations
and M. Markevitch with information about the ACIS background.
This work was supported in part by NASA grant GO7-8127X.

\end{document}